\documentclass[aps,pre,twocolumn,showpacs,amsfonts,floatfix,nofootinbib,superscriptaddress]{revtex4-1}

\usepackage{graphicx,epstopdf}              
\usepackage{subcaption}						
\usepackage{hyperref}                   	
\usepackage{amsmath,amssymb}               	

\graphicspath{{figs/}}						

\usepackage{color}
\usepackage{dcolumn}
\definecolor{OliveGreen}{RGB}{0,102,0}

\hypersetup{colorlinks=true,urlcolor=blue,linkcolor=blue,citecolor=red}

 \providecommand{\abs}[1]{\vert #1\vert}

\newcommand{\lrangle}[1]{\langle{#1}\rangle}

\begin{document}
\title{Universality and crossover behavior of single-step growth models in $1+1$ and $2+1$ dimensions}
\author{E. Daryaei}
\affiliation{Department of Physics, Faculty of Basic Sciences, University of Neyshabur, P.O. Box 91136-899, Neyshabur, Iran} 

\begin{abstract}
We study the kinetic roughening of the single-step (SS) growth model with a tunable parameter $p$ in $1+1$ and $2+1$ dimensions by performing extensive numerical simulations. We show that there exists a very slow crossover from an intermediate regime dominated by the Edwards-Wilkinson class to an asymptotic regime dominated by the Kardar-Parisi-Zhang (KPZ) class for any $p <\frac{1}{2}$.
We also identify the crossover time, the nonlinear coupling constant, and some nonuniversal parameters in the KPZ equation as a function $p$.
The effective nonuniversal parameters are continuously decreasing with $p$, but not in a linear fashion. Our results provide complete and conclusive evidence that the SS model for $p \neq \frac{1}{2}$ belongs to the KPZ universality class in $2+1$ dimensions.

\end{abstract}

\maketitle
 
\section{Introduction}
Understanding the kinetic roughening of growing surfaces and interfaces has attracted much interest from both theoretical and experimental points of view~\cite{Barabasi95,Krug97,*Halpin-Healy95}. Since four decades ago, a dynamic scaling approach was proposed to describe the morphological evolution of a growth front and various discrete models have been suggested to describe surface growth processes, for example see~\cite{Barabasi95,Meakin97}. These discrete models can be described by some continuous Langevin equations. Two well-known Langevin equations are the Edwards-Wilkinson (EW)~\cite{Edwards82} and the Kardar-Parisi-Zhang (KPZ)~\cite{Kardar86} equations. A large class of discrete growth models such as the ballistic deposition (BD) models~\cite{Marjorie59}, restricted solid on solid (RSOS) models~\cite{Kim89}, and directed polymers in random media~\cite{Kardar87} are believed to belong to the same universality class as the KPZ equation describing the growth interface fluctuations.  
The KPZ equation describes the time evolution of a field
$h(\mathbf{x},t)$ that denotes its height at the position $\mathbf{x}$ and at time $t$ on a $d-$dimensional substrate: 
\begin{equation}
\frac{\partial h(\mathbf{x}, t)}{\partial t}= \nu\,\nabla^2 h  + \frac{\lambda}{2}\,\abs{\nabla h}^2 + \sqrt{D} \, \xi(\mathbf{x},t) \;,
\label{eq:KPZ}
\end{equation}
\noindent where $\xi(x,t)$ is an uncorrelated Gaussian white noise in both space and time with
zero average {\it i.e.} $ \langle \xi(\mathbf{x},t) \rangle = 0 $ and $  \langle \xi(\mathbf{x},t)\xi(\mathbf{x'},t') \rangle  = \delta^d (\mathbf{x}-\mathbf{x'}) \delta(t-t')$. The real constants $\nu$, $\lambda$, and $D$ take into account the surface relaxation intensity, the lateral growth and the amplitude of Gaussian white noise, respectively. One of the most important quantities that can be used to study and to classify different discrete or continuous growth models, like Eq.~(\ref{eq:KPZ}), is defined in 
terms of the scaling properties of the surface width $w(L,t)
= \sqrt{ \langle [ h(\mathbf{x},t)- \langle h(\mathbf{x},t) \rangle ]^2\rangle }$ where $ \langle \cdot \cdot \rangle $ denotes average over space and ensemble realizations. As a function of the system size $L$, it is expected to have the scaling form
$w(L,t) \sim L^{\alpha}f(t/L^z)$~\cite{Family85}, where  $\alpha$ and $z$ are two independent universal parameters known as roughness and dynamic exponents, respectively. The scaling function $ f $ usually has the asymptotic form $f(x\gg1)=\mathrm{constant}$ and $f(x\ll1)\sim x^{\beta}$, where $\beta$ is the growth exponent $\beta=\alpha/z$. The particular behaviors
of $f$ imply that $w(L,t) \sim L^{\alpha}$ for $t \gg L^z$ and
$w(L,t)\sim t^{\beta}$ for $t \ll L^z$.
The absence of the nonlinear term, {\it i.e.} Eq.~(\ref{eq:KPZ}) with $\lambda=0 $, results in another universality class known as the EW where the exact values of exponents are given by $ \alpha = (2-d)/2 $  and $ z = 2 $, in $(d+1)$ dimensions~\cite{Edwards82}.
In the presence of $\lambda$, although, due to the Galilean invariance, another scaling relation $\alpha+ z = 2$ emerges \cite{Forster77}, the exact solution only exists in $d=1+1$ which gives $ \alpha=1/2 $, and $z=3/2$~\cite{Kardar86}. In higher dimensions, the critical exponents are available only by various theoretical approaches \cite{Frey94,*Moshe92,*Kloss12} and numerical methods~\cite{Pagnani16,Kelling16,Kelling17}.

In the breakthrough theoretical approach \cite{Johansson00}, Johansson successfully computed a universal probability distribution function (PDF) for a discrete growth model, known as single-step (SS)~\cite{Meakin86, Plischke87, Liu88, Kondev00}. Most especially, the PDF of the height fluctuations is the {\em Tracy-Widom} (TW) distribution \cite{Tracy94}, which in the context of the random matrix theory, describes the typical fluctuations of the largest eigenvalue of random matrices belonging to the Gaussian Unitary Ensemble (GUE) \cite{Forrester93}.
In $d=1+1$, the surface (or interface) height in the KPZ systems asymptotically evolves according to the {\it ansatz}~\cite{Krug92,Johansson00,Prahofer00}
\begin{equation} 
h \simeq v_{\infty} t + s_{\lambda} (\Gamma t)^{\beta} \chi, 
\label{eq::ansatz} 
\end{equation}
\noindent where $\chi$ is a stochastic variable that carries universal information of the fluctuations, while the system-dependent parameters $v_{\infty}$, $s_{\lambda}$, and $\Gamma$ are the asymptotic interface velocity, the signal of $\lambda$ in the KPZ equation Eq.~(\ref{eq:KPZ}), a non-universal constant associated with the amplitude of the interface fluctuations, respectively. 
Remarkably, there are a few non-Gaussian universal distributions that $\chi$ selects one of them based on the global geometric shape of the initial condition $h(\mathbf{x},t=0)$, namely the Gaussian orthogonal ensemble (GOE) distributions for initial flat interfaces and the GUE distributions for curved ones \cite{Calabrese10,Dotsenko10,Sasamoto10, Amir11}. This geometry-dependent universality was tested and confirmed experimentally, in studies on growing interfaces of nematic liquid crystals
 \cite{Takeuchi10,*Takeuchi11}. Recent numerical simulations have shown that the KPZ {\it ansatz}, {\it i.e.} Eq.~(\ref{eq::ansatz}), can be generalized to 
$2+1$ dimensions~\cite{Oliveira13,Halpin12,Halpin13}, but the exact forms of the asymptotic distributions of $\chi$ are yet not known.

Although the first studies of the TW fluctuations were initially performed on the SS model in $(1+1)$ dimensions and for a wedge initial condition~\cite{Johansson00}, numerical simulations in higher dimensions, except for a few reports \cite{Oliveira13,Carrasco14} under a particular condition, commonly fail to provide a reliable connection between this model and the KPZ class. Moreover, this model can be mapped onto some extensively studied models in equilibrium or nonequilibrium statistical mechanics, such as the kinetic Ising model \cite{Meakin86,Plischke87}, the asymmetric simple exclusion process \cite{Derrida98}, and the six-vertex model \cite{Meakin86,Neergaard97,Gwa92}. Therefore, some properties of the SS model can be acquired analytically from the exact results of these well-studied models \cite{Barabasi95, Neergaard97,Plischke87}. 
In this paper, we study the SS model, which is defined in the following way: at any time $t$, we randomly select a site $i$ on the $d-$dimensional lattice, and we let the surface height $h_i$ at that site to increase by 2 with probability $p$ only if it is a local minimum, or to decrease by 2 with probability $q$ only if it is a local maximum. For simplicity, and without any loss of generality, we can impose $q=1-p$ condition.
Since the height difference between two neighboring sites can only be two values (+1 or -1), the SS model is analytically more tractable \cite{Neergaard97,Johansson00}.

In $1+1$ dimensions, it is known that this model can be exactly solved by mapping to the kinetic Ising model~\cite{Plischke87,Barabasi95}, and belongs to EW (KPZ) universality class for $ p=0.5 $ ($ p \ne 0.5 $) ~\cite{Plischke87, Liu88, Kondev00}.
In contrast to the deep understanding of the SS model in $d=1+1$, essentially conflicting results still exist regarding the scaling behaviors of this model in $d =2+1$.
Although it is generally agreed upon that the SS model belongs to the KPZ universality class for $p=0$ and the EW class for $p=0.5$, the probability interval in which the model is consistent with the EW or KPZ classes is a matter of contention.
In some reports, the nonlinearity coefficient $\lambda$ in the KPZ Eq.~(\ref{eq:KPZ}) have been considered as proportional to $p^{\prime}\equiv(q-p)$ and concluded that the model asymptotically belongs to the KPZ universality class for all $p\ne 0$.\cite{Liu88,Plischke87,Tang90,Tang92,Forrest90}. However, some authors \cite{Kondev00,Dashti17} found that there exists a critical value $p_c$ around which for $p>p_c$ the model consistently resembles $p=0.5$.
More recently, a geometrical investigation \cite{Dashti17} reported a roughening transition around $p_c\approx 0.25$ from a rough phase in the KPZ universality to the smooth phase in the EW universality class.\\
In this paper, we revisit the kinetic roughening of the SS model in $d=2+1$ 
to address the seeming disagreement between the studies more thoroughly and to perform careful finite-size and finite-time analysis. Using extensive numerical simulations, as we will see in the following, we show that there exists a slow crossover from an intermediate regime dominated by the EW class to an asymptotic regime dominated by the KPZ class for any $p^{\prime}\neq 0$. Therefore, our results rule out any roughening transition in $2+1$ dimensions \cite{Dashti17}. Additionally, we are going to estimate all universal and nonuniversal
parameters related to the KPZ Eq.~(\ref{eq:KPZ}) and the KPZ {\it ansatz} given by Eq.~(\ref{eq::ansatz}).
We also consider the SS model in 1+1 dimensions, since there the universal and the nonuniversal parameters, as well as asymptotic behavior of this model are well known and this, therefore, provides a convenient test for our numerical results.

The rest of the paper is organized as follows. The simulation details are presented in Sec.~\ref{Sec::simulation_details}. The scaling behaviors of surface width and related consequences are discussed in Sec.~\ref{Sec::surface_width}. The interface velocities of SS model are estimated for different values of $p$ in Sec.~\ref{Sec::interface-velocity}, and in the following, the nonuniversal parameters in the KPZ {\it ansatz} given by Eq.~(\ref{eq::ansatz}) as a function of the control parameter $p$ are determined in section \ref{Sec::non-universal}. Final discussions and conclusions of the SS model are presented in~Sec.\ref{Sec::Conc}.

\begin{figure}[!]
\begin{center}
\begin{tabular}{lll}
\includegraphics[scale=0.22]{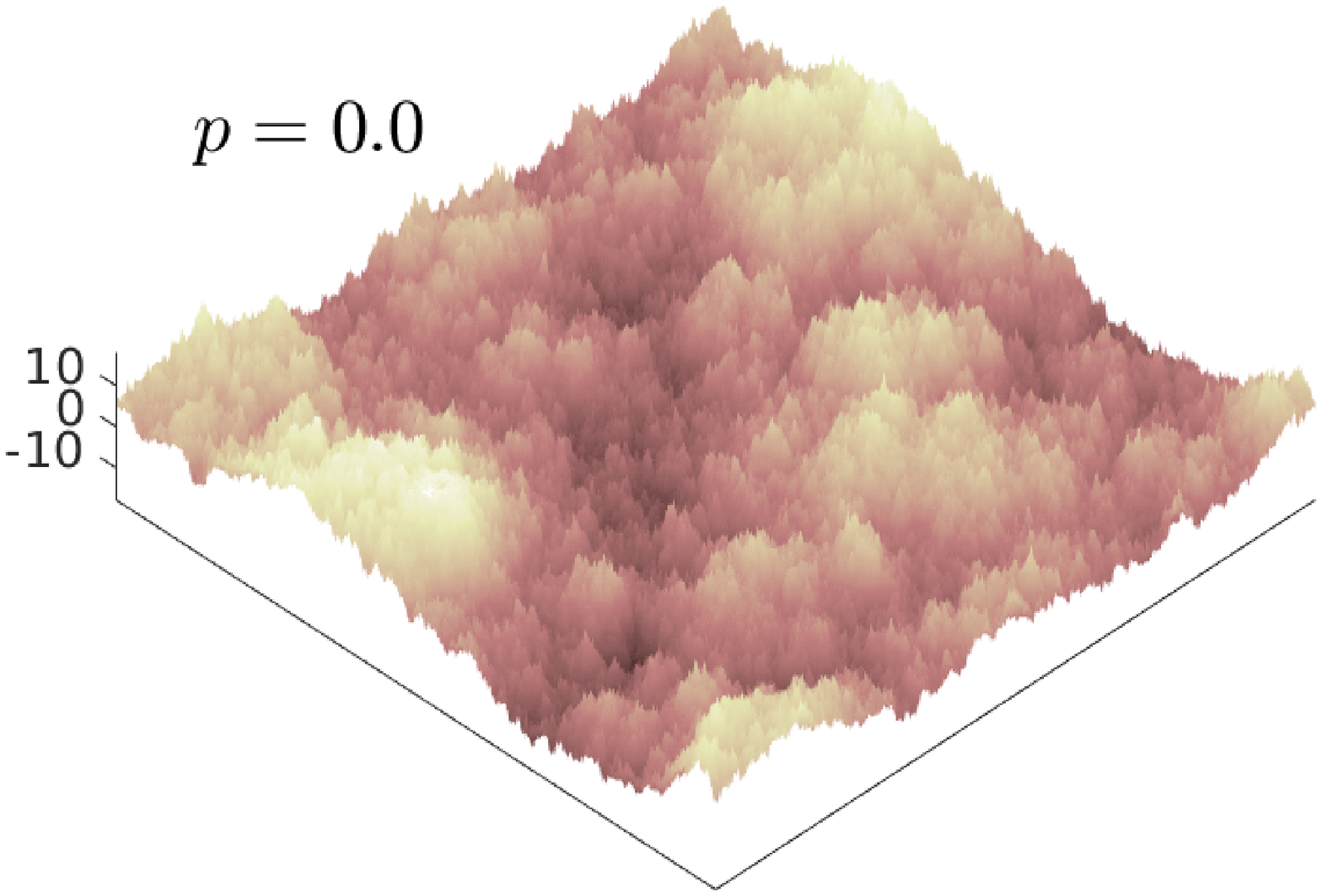}
 \includegraphics[scale=0.22]{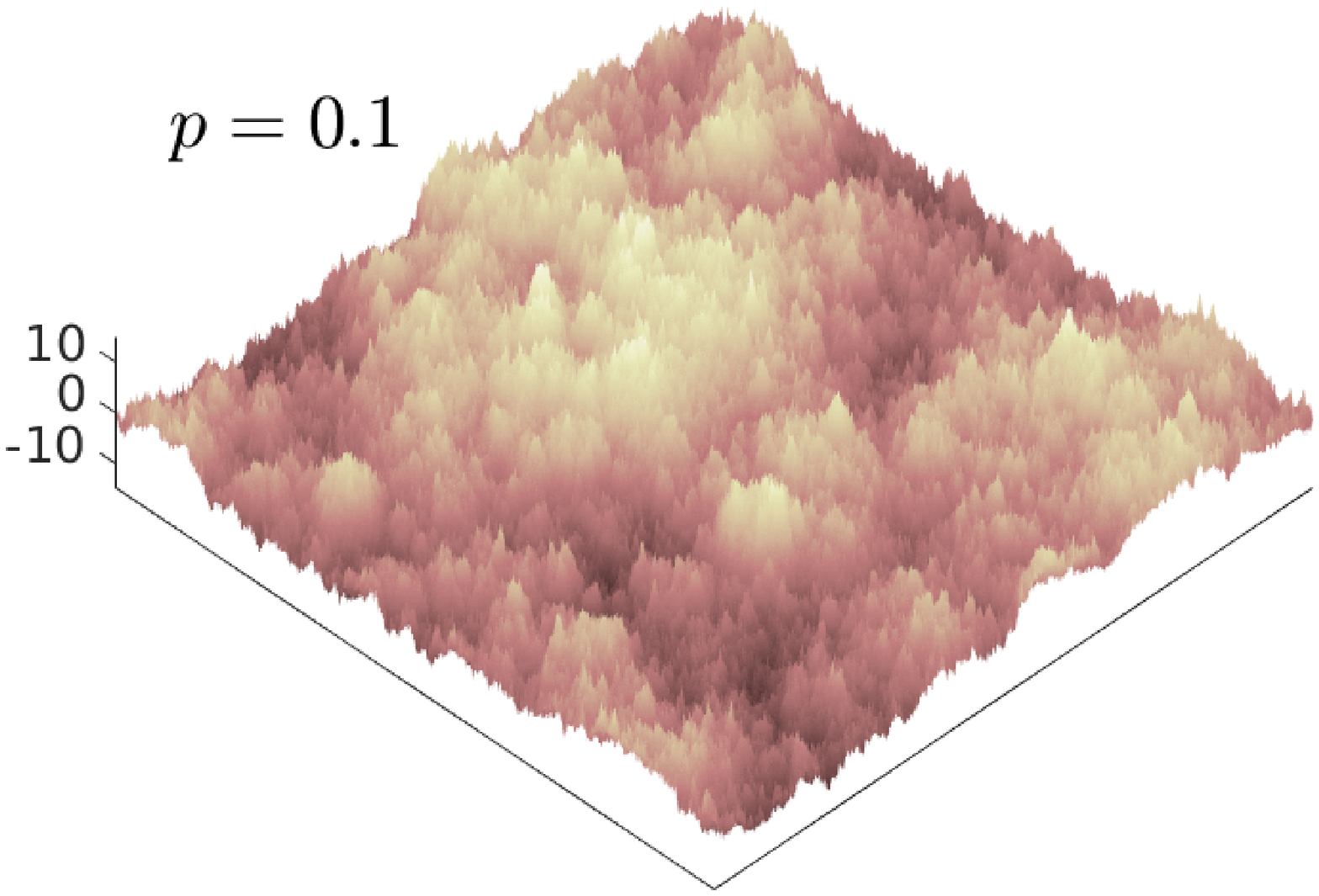}
 \end{tabular}
 \begin{tabular}{lll}
 \includegraphics[scale=0.22]{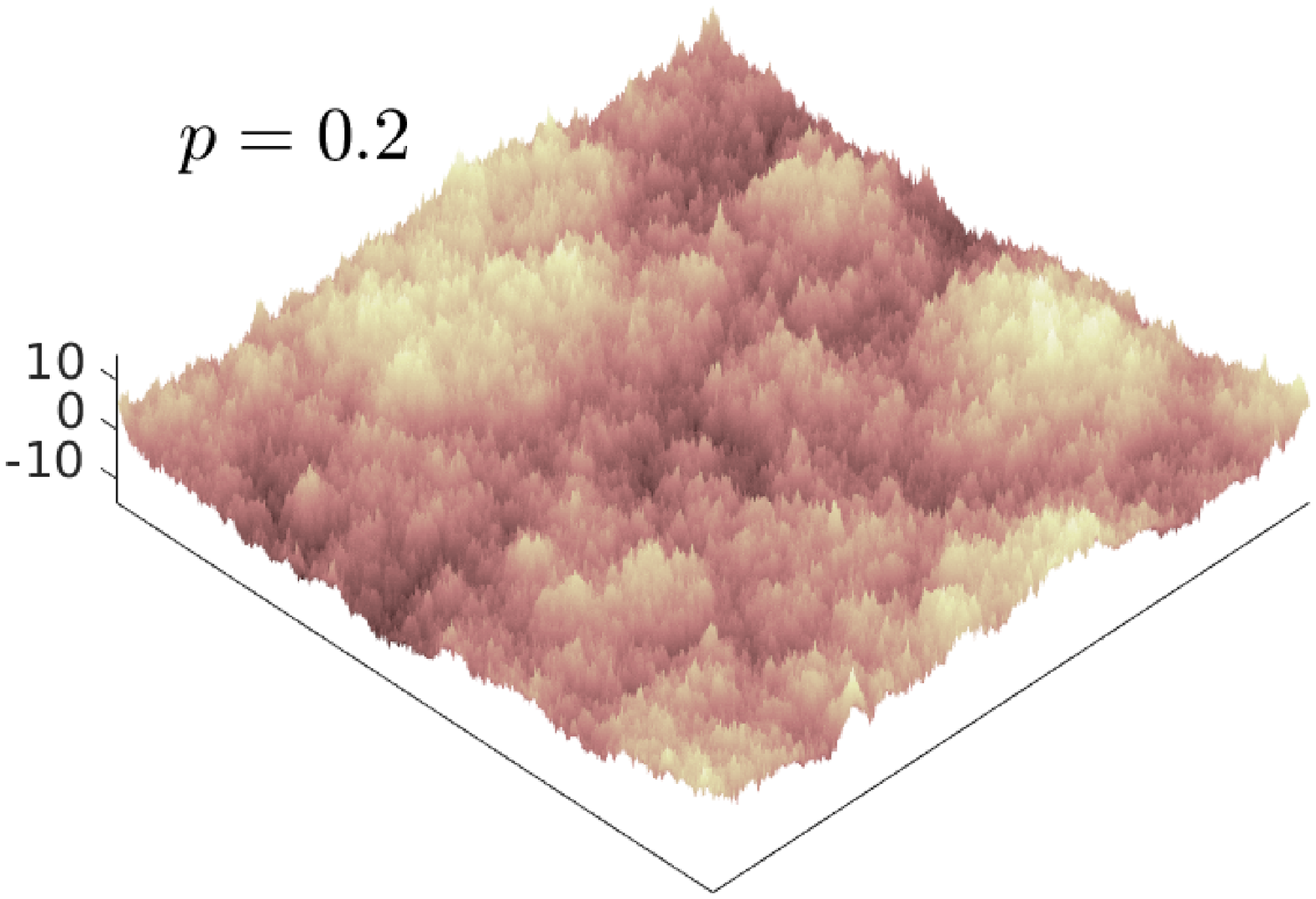}
 \includegraphics[scale=0.22]{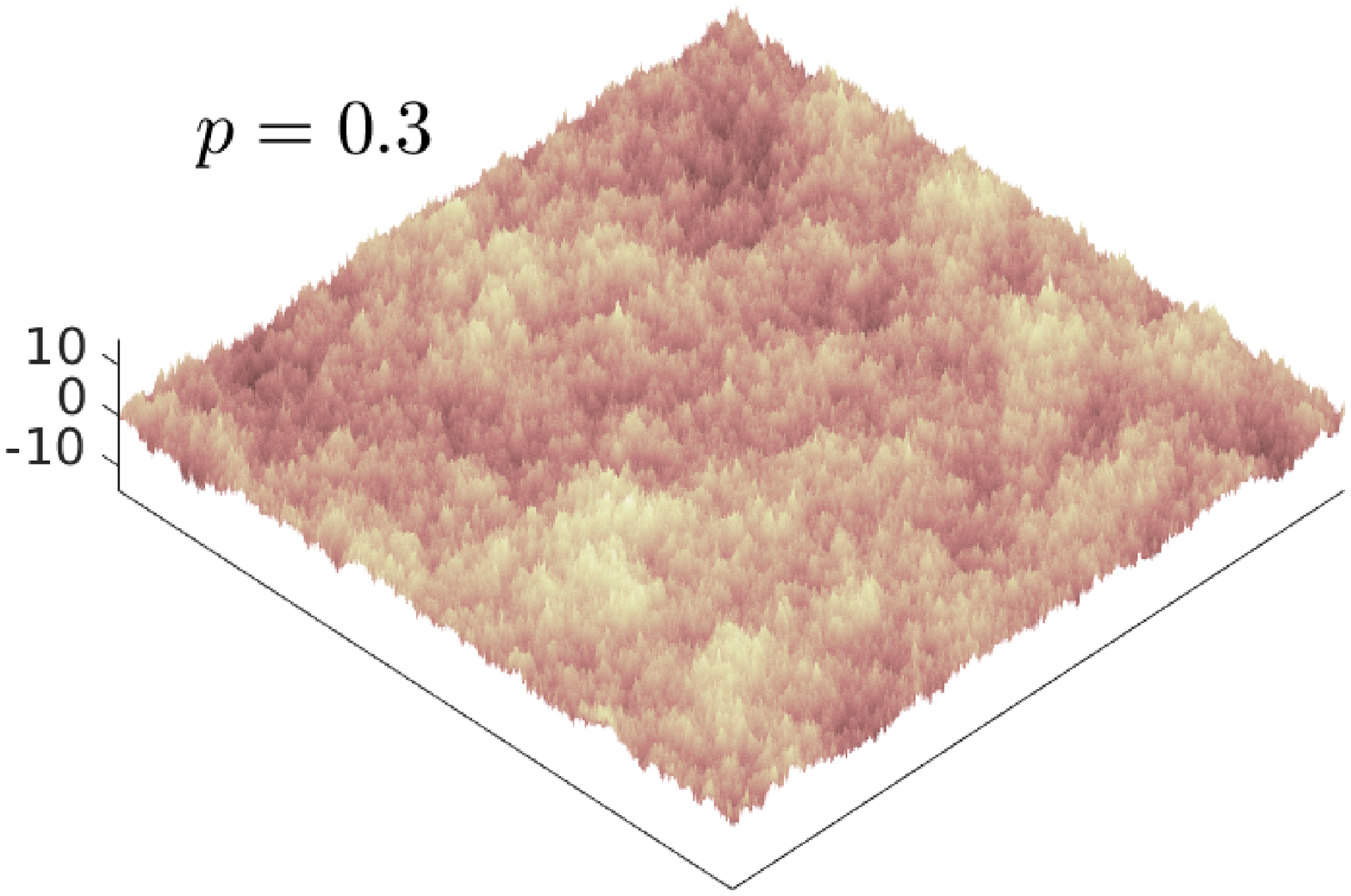} 
 \end{tabular}
\begin{tabular}{lll}
 \includegraphics[scale=0.22]{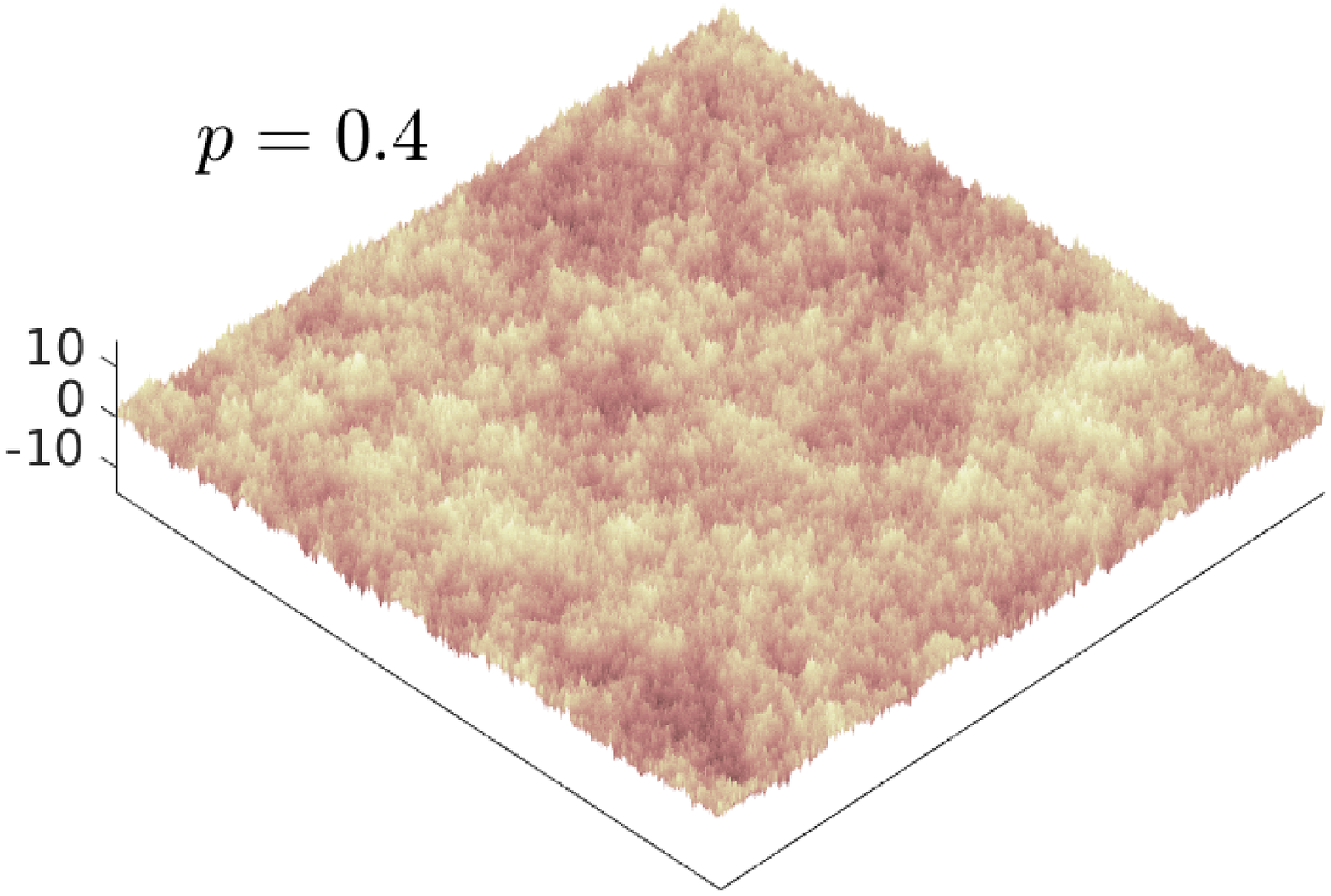} 
 \includegraphics[scale=0.22]{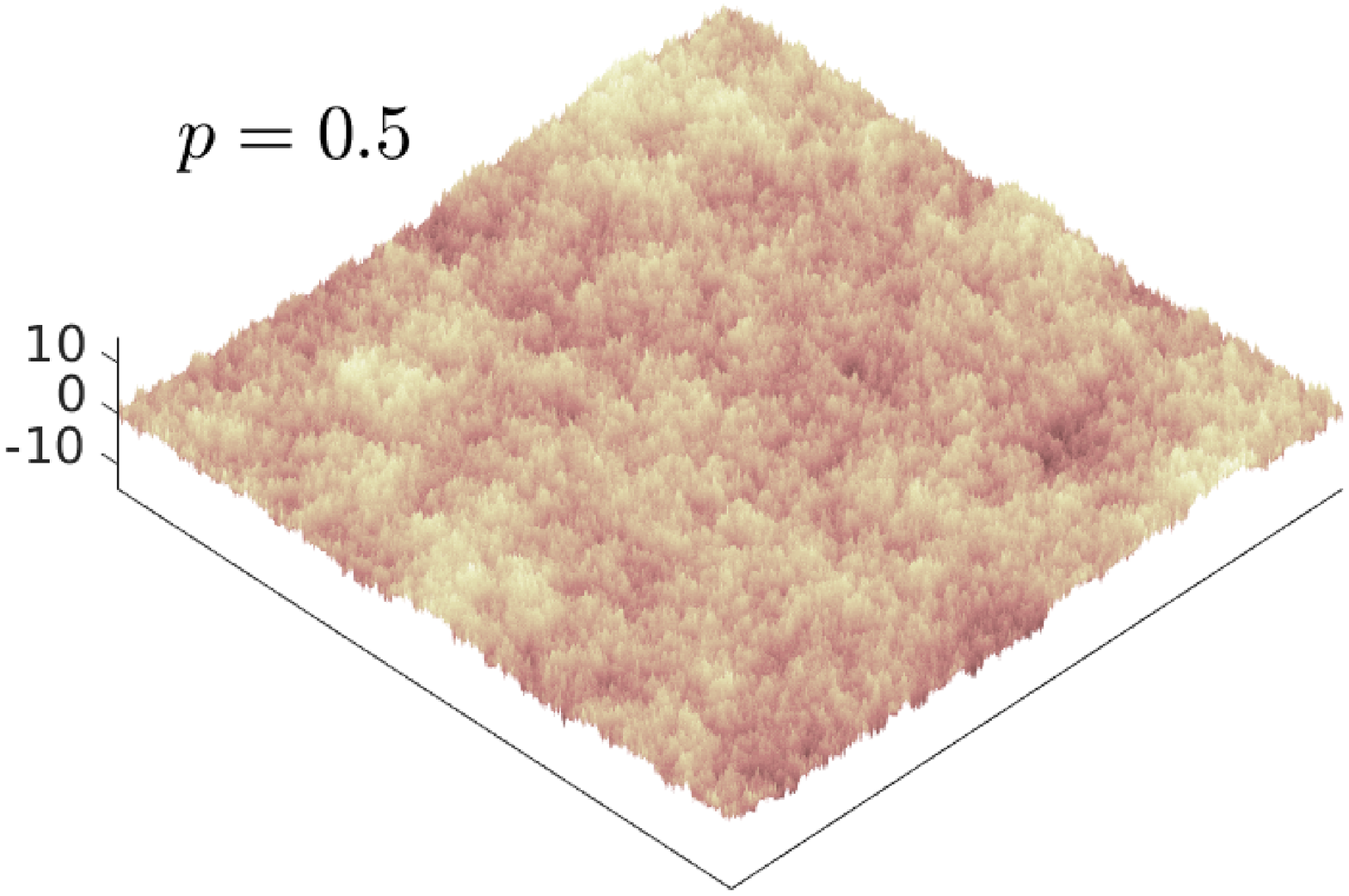}
\end{tabular}
 \caption{(Color online) Snapshots of a typical surface morphology grown by SS model on a two-dimensional lattice of size $1024$ at the steady-state regime for various values of $p$.}
\label{fig:surface-Snapshots}
\end{center}
\end{figure}

\section{Simulation details}
\label{Sec::simulation_details}
We performed extensive simulations of the SS model on two-dimensional lattices of size \mbox{$L=2^{n+3},$} \mbox{$n=1, 2, ..., 8$} with periodic boundary conditions. Throughout this study, we used the implementation of the sequential updating rule described above. The number of samples generated for each lattice size ranges from $5\times10^5$ for the smallest lattice sizes till about $200$ for the largest lattice sizes. Moreover, size $ L=2500 $ was only used to inspect the scaling behavior of the SS model at $p=0.25$. A checkerboard initial condition, as described in~\cite{Dashti17}, has been used. Moreover, to observe the crossover behavior, and to check our algorithms with exact analytic results, we simulated $(1+1)$-dimensional SS model up to size $2^{17}$. We also simulated the BD model and numerically obtained $v_{\infty}$, $\lambda$, and $\Gamma$ and finally checked them with more accurate results \cite{Alves14}. In our numerical simulation of the SS model, we impose the condition $p+q=1$, so due to up/down symmetry in our model definition, we just need to consider $p \le 0.5$. Throughout this paper, the time is measured by Monte Carlo steps per site.
The surface morphology grown by the KPZ equation is characterized by relatively large hills and deep valleys that lead to rough surface morphology, while the EW equation produces a very smooth surface and the size of the hills is negligible in comparison to the lattice size. To observe these morphological differences, we simulate a few samples on a lattice of size $1024$, for different values of $p$.
The surface morphologies of $(2+1)$-dimensional SS model for various values of $p$ are shown in Fig.~\ref{fig:surface-Snapshots}. As expected, the surface morphology decreases with $p$. At first glance, one would find a smooth surface on higher values of $p$ but, in principle, as we will see in the following, this can be described as a result of finite-size effects.

\section{Surface width and crossover behavior}
\label{Sec::surface_width}

\begin{figure}
\centering
\includegraphics[width=0.44\textwidth]{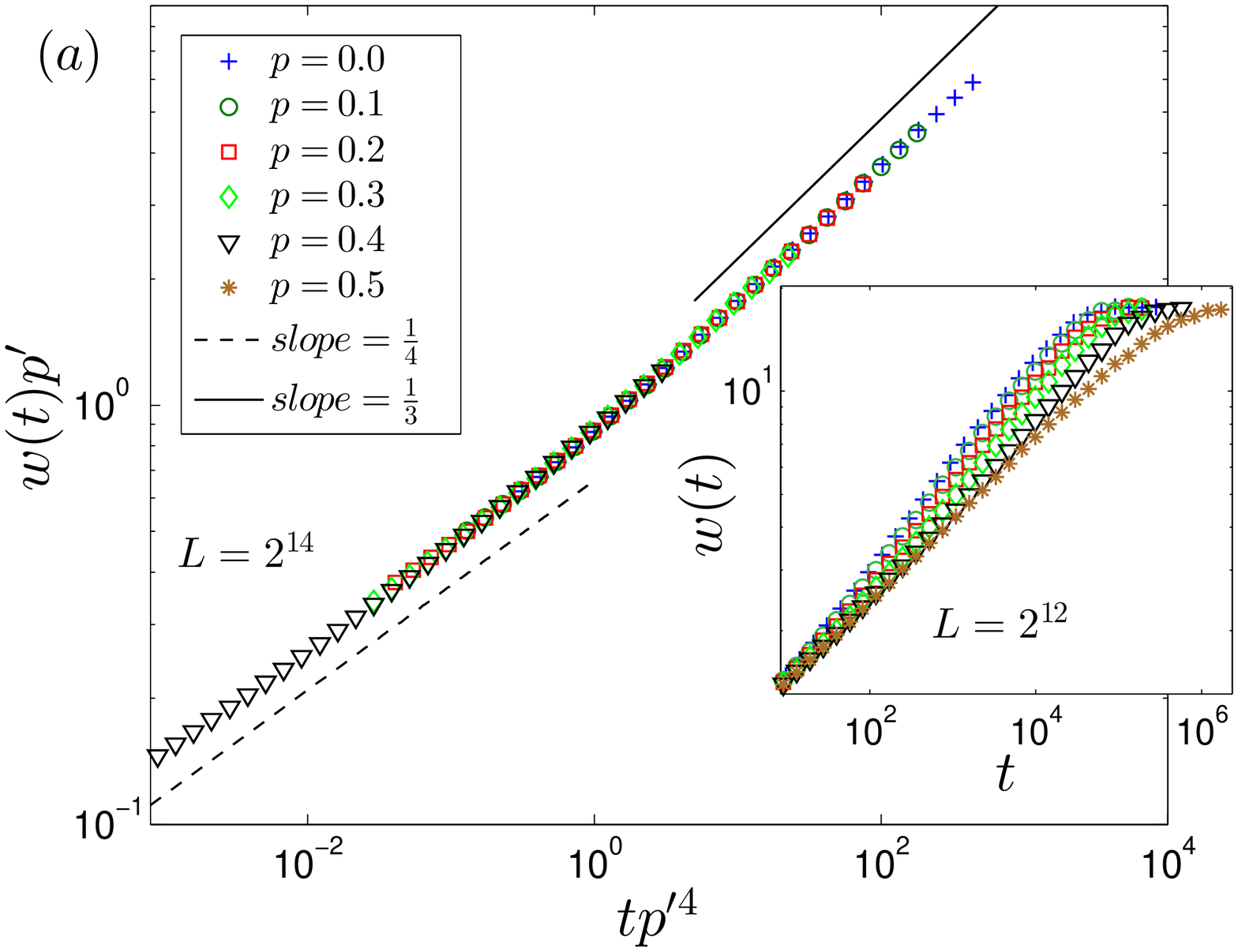} 
\includegraphics[width=0.48\textwidth]{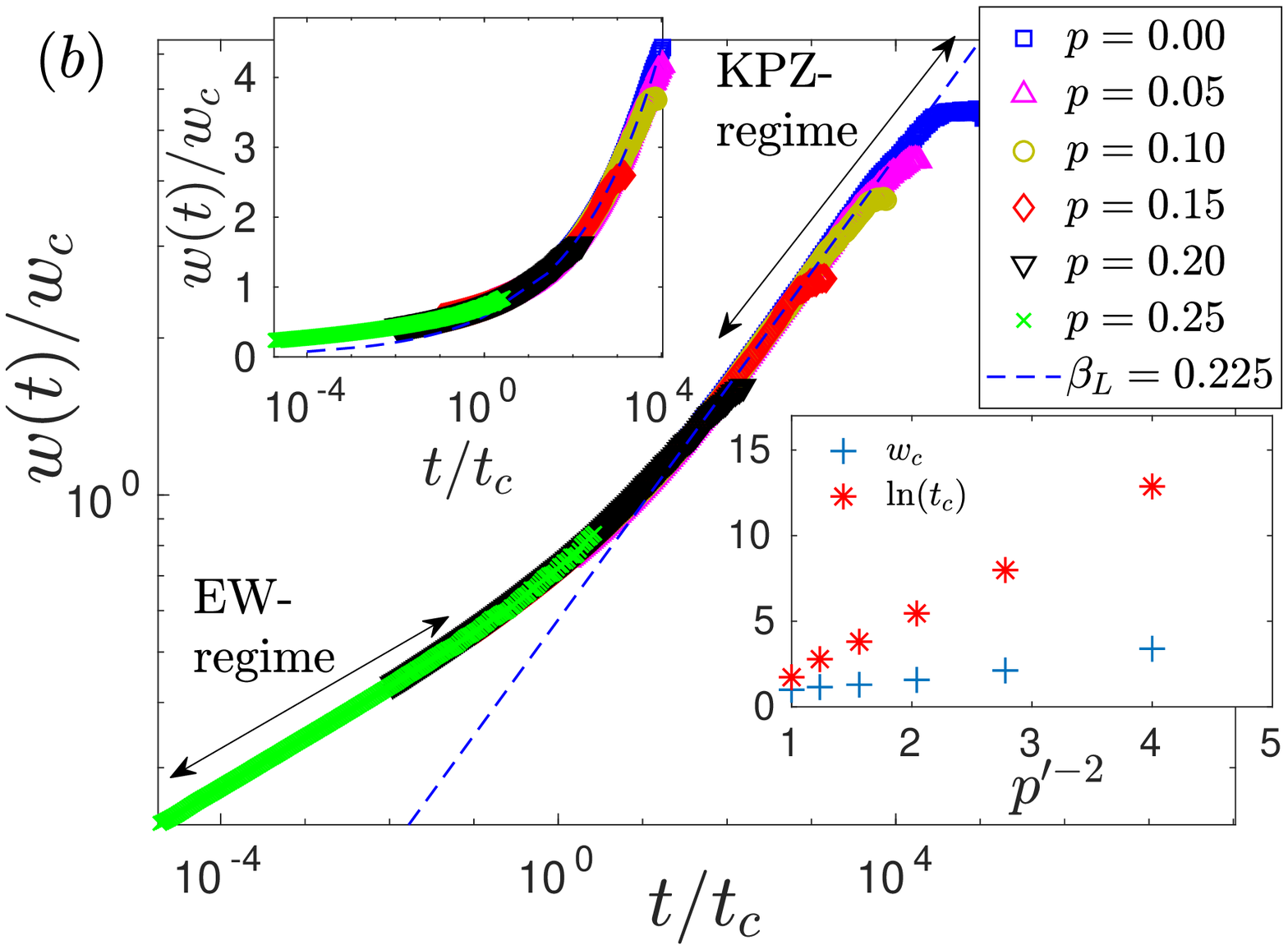}
\caption{(Color online) Scaling plots of the rescaled interface width $w/w_c$ vs the rescaled time $t/t_c$ in $d=1+1$ \textbf{(a)}, and $d=2+1$ \textbf{(b)}. The lattice size for $(2+1)$-dimensional case is $2^{10}$ (except for $p=0.25$ which used a lattice of size 2500). The dashed line indicates a power-law fitted curve with effective growth exponent $\beta_L=0.225$ for $p=0$ on the growth regime. Inset of (a): The full time on a lattice size of $2^{12}$. Upper-Inset of (b): Same data in a semi-log plot. Lower-Inset of (b): $w_c$ and $\ln(t_c)$ against $p'^{-2}$.} 
\label{fig::surface_w}
\end{figure}
It is known that in the short time limit, the non-linear term in Eq.~(\ref{eq:KPZ}) is less important than the linear Laplacian term. The typical surface width, in this limit, is well described by the EW equation. In fact, in $2+1$ dimensions and below, depending on the non-universal parameters, both discrete and continuous growth models present a crossover time $t_c$. As a matter of fact, to observe this crossover, the system size must be large enough so that saturation effects take place much later than the crossover time $t_c$ {\it i.e.} $L^z \gg t_c$. Therefore, the minimum system size required to occur this crossover behavior $l_{c}$, approximately scales as $t_c^{1/z}$.

In $d=1+1$, for simplicity we can work in rescaled units: $x \to l_cx$, $t \to t_ct$, and $h \to h_ch$ where from dimensional analysis these characteristic scales of space, time, and height can be obtained as \cite{Amar90,Sneppen92} 
\begin{align}
\centering
 l_c&=\frac{(2 \nu)^3}{D \lambda^2} ~, &  t_c&=2\frac{(2 \nu)^5}{D^2 \lambda^4} ~,  & h_c&=\frac{2 \nu}{\lambda}
 \label{eq:rescaled_units}
\end{align}
The crossover time $t_c$, and local surface height at crossover point $h_c$, as well as the crossover surface width $w_c$ scale as $\nu^5/(D^2 \lambda^4)$, and $\nu/\lambda$ respectively.
For the $(1+1)$-dimensional SS model, the exact analytic result for the coefficient of the nonlinear term in the KPZ equation is known as $\lambda =(q-p)$~\cite{Barabasi95,Krug92}.
The strength of the white noise $D$ is kept constant and independent of $p$ during the simulations \footnote{This is due to the fact that, as will be discussed later, in the ($1+1$)-dimensional SS model the probability of choosing a site eligible for growth (or desorption) is independent of the value of $p$.}.
Additionally, the $D/\nu$ ratio in the Eq.~(\ref{eq:rescaled_units}) is related to the {\it steady-state} width of the interface, which scales with the finite system size L via the relation $w_{sat}\sim \sqrt{LD/\nu}$~\cite{Krug92}. As shown in the inset of Fig.~\ref{fig::surface_w}(a), the saturated surface width is independent of the value of the parameter $p$, consequently, we expect that the surface relaxation intensity, $\nu$ is independent of $p$. Thus, the $\lambda$ parameter is responsible for the variation of the $t_c$ and $w_c$. 
As $p$ increases towards $0.5$, based on a naive scaling analysis of Eq.~(\ref{eq:rescaled_units}), the crossover time $t_c$ and the crossover surface width $w_c$  diverge as $p^{\prime-4}$ and  $p^{\prime-1}$, respectively.
However, to confirm this prediction, as shown in Fig.~\ref{fig::surface_w}(a), we plot the rescaled interface width $wp^{\prime}$ as a function of the rescaled time $tp^{\prime 4}$ which is in excellent agreement with the analytic predictions. Such dependence has been observed in some competitive models~\cite{Oliveira06,Tang92}.

Since $d=2+1$ is the marginal dimension \cite{Barabasi95,Tang90,Kardar86}, we cannot follow the dimensional analysis approach. Based on RG analysis, it is known that the crossover length scale $l_c$ displays an exponential dependence on the value of the effective coupling constant $g\sim\frac{D\lambda^2}{\nu^3}$ \cite{Tang90}. In the absence of exact analytical results for $t_c$ and $w_c$ as a function of $p$ for the SS model, we can numerically estimate $t_c$, and $w_c$, by rescaling the time $t$ and surface width $w$ by arbitrary values for $t_c$, and $w_c$, respectively, to have a good data collapse, as shown in the Fig.~\ref{fig::surface_w}(b). 
The surface width cross from an intermediate regime dominated by the EW regime (logarithmic-law) to an asymptotic regime dominated by the KPZ regime (power-law). Our observation is in agreement with the slow crossover scenario discussed in \cite{Tang90,Tang92}. By increasing the value of $p$, the crossover time $t_c$, as well as crossover length $l_c$, increase exponentially. For example, to observe the crossover behavior for $p=0.25$, we need a lattice of size around $2500$, which after a typical time $4\times10^5$, it arises. 

To verify that $t_c$ increase exponentially with $p$, we try to find a linear dependence between $\ln(t_c)$ and an approximate function of $p'^{-1}$.
As shown in the inset of Fig.~\ref{fig::surface_w}(b), we can find a good linear dependence of $\ln(t_c)$ on the $p'^{-2}$ parameter which was reported in a similar growth model with slow crossover in \cite{Tang92}.
It is also worth mentioning, for the values $p>0.25$, due to this slow crossover, we are not able to observe the crossover behavior in a reasonable amount of computational time. For instance, for $p=0.3$, based on an extrapolation method, we estimate its crossover time  approximately $3\times10^8$. So, to observe this crossover, the lattice size must be large enough so that saturation effects take place much later than this time.

\begin{figure}[!]
\centering
\includegraphics[width=0.49\textwidth]{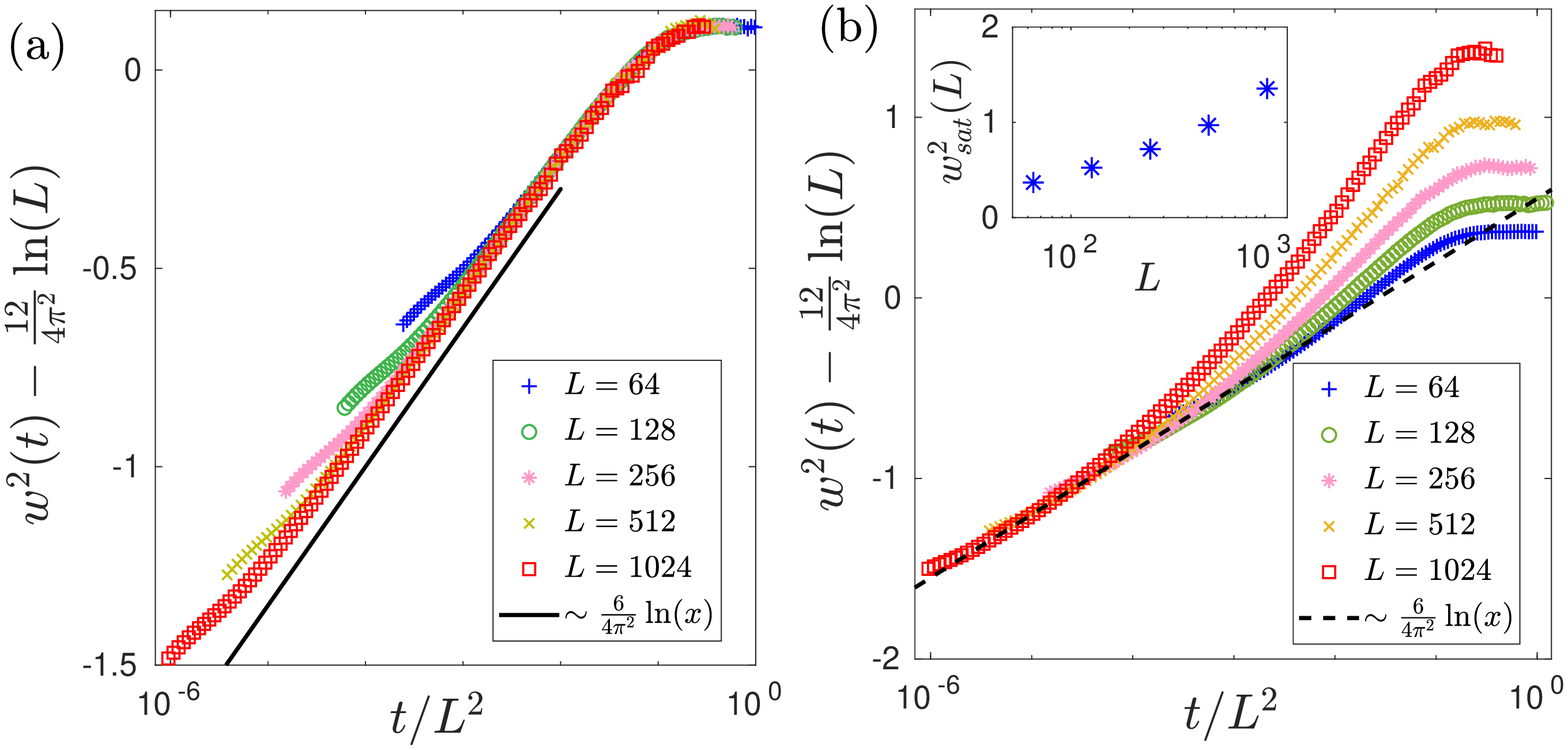}
\includegraphics[width=0.48\textwidth]{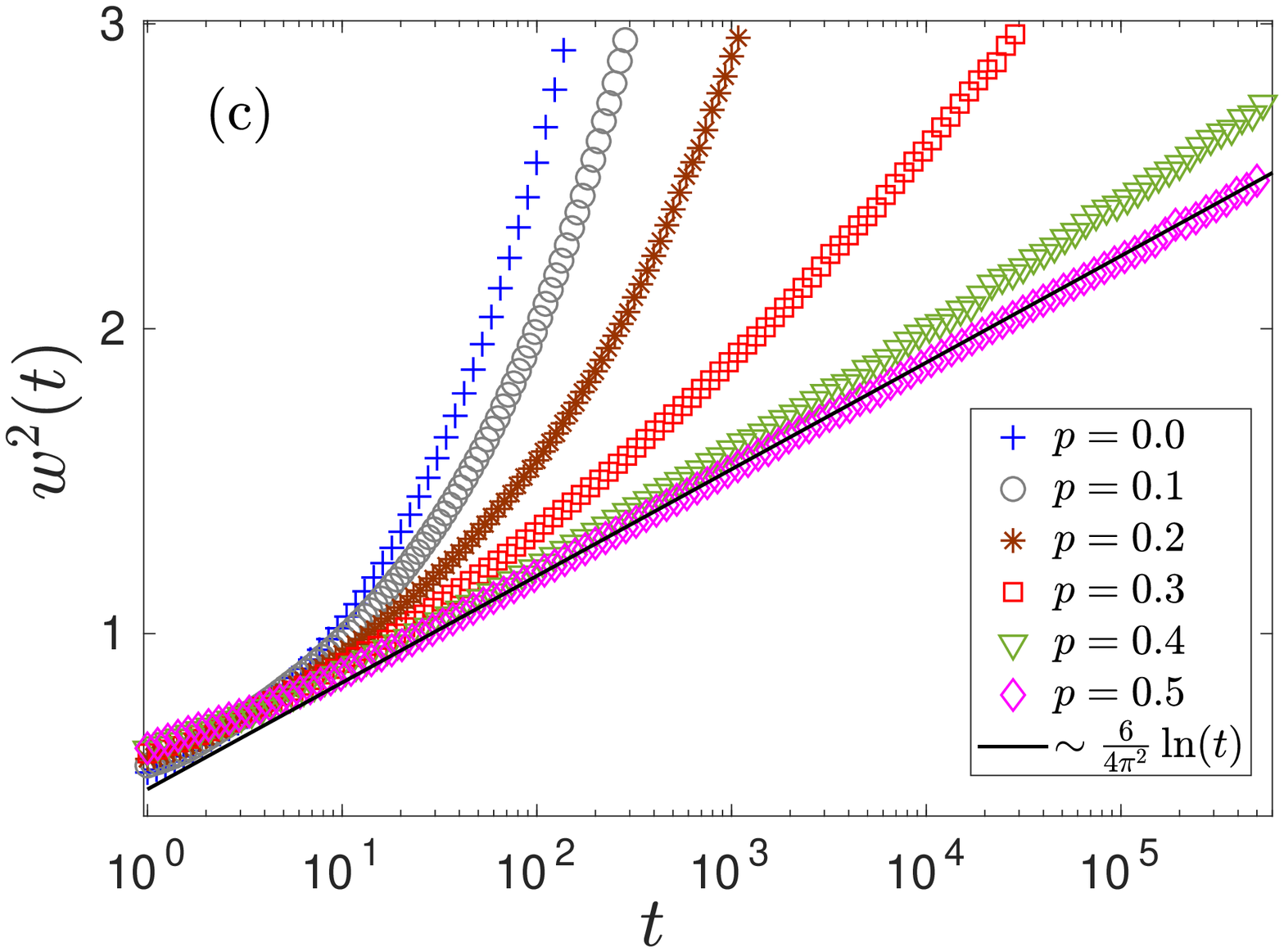}
\caption{(Color online) Examination of logarithmic surface growth for $(2+1)$-dimensional SS model.
Scaling plot of the rescaled $w^2(t)-2\delta \ln(L)$ at $p=0.5$ \textbf{(a)}, and $p=0.3$ \textbf{(b)} versus the rescaled time $t/L^2$ for different lattice sizes. Although we observe an excellent data collapse at $p=0.5$ for $\delta=\frac{6}{4\pi^2}$, we are not able to find a good data collapsing at $p=0.3$ for any prefactor of $\delta$. The inset of (b) shows the saturated width $w^2_{sat}$ versus the system size $L$ in a semilogarithmic scale. \textbf{(c)} Variation of $w^2(t)$ vs time $t$ on the initial growth regime for different probabilities $p$ by considering a system size $L= 2500$.
}
\label{fig::initiall}
\end{figure}
 
In the case of up-down symmetry, {\it i.e.} $p=0.5$, the nonlinear term
is dropped, and the KPZ Eq.~(\ref{eq:KPZ}) simplifies to the EW equation \cite{Edwards82}.
In $d=2+1$, since the growth and the roughness exponents of the EW equation vanish ({\it i.e.} $\beta=\alpha=0$), a logarithmic scaling arises from mean-field theory.
Indeed, the surface width grows like
$ w^2(t,L) \sim  \delta \ln (t) $ in initial growth time, {\it i.e.} $ t \ll L^2 $, and becomes saturated like $ w^2(t,L) \sim  2 \delta \ln (L) $ in the stationary regime, {\it i.e.} $ t \gg L^2 $.
The prefactor $\delta$ is exactly known for the EW equation $D/(4\pi\nu)$ \cite{Nattermann92}. By taking into account the exact value of surface tension in a triangular Ising antiferromagnet system \cite{Blote82}, and by using some geometrical transformations for a driven lattice gas model of dimers \cite{Odor09} which is similar to the SS model in the case of $p=0.5$, the theoretical estimation results in $\delta=\frac{6}{4\pi^2} \simeq 0.15198$ \cite{Odor09}.
To confirm this prediction for $p=0.5$, as shown in the Fig.~\ref{fig::initiall}(a), we plot the rescaled surface width $w^2(t)-\frac{12}{4\pi^2} \ln(L)$ as
a function of the rescaled time $t/L^2$ in a semilogarithmic scale. As can be seen, a very good data collapse is obtained which is in excellent
agreement with the analytic predictions.

To inspect the validity of the roughening transition for $p<0.5$, we also plot the rescaled surface width at $p=0.3$ versus the rescaled time for different lattice sizes in Fig.~\ref{fig::initiall}(b). As can be seen, we are not able to observe a good data collapsing at $p = 0.3$ for any prefactor of $\delta$. If we set the exact theoretical value for $\delta$, we observe a good data collapse, in particular, just for the initial stage of the growth. Moreover, as shown in the inset of Fig.~\ref{fig::initiall}(b), the saturation roughness distinctly increases faster than $\log(L)$, which is another evidence of the crossover to KPZ class.
To see how $(2+1)$-dimensional SS model behaves in the initial growth time, and how it violate the linear behavior of $w^2(t)$ in a semilogarithmic scale, in the Fig.~\ref{fig::initiall}(c), we also plot, $w^2(t)$ for different probabilities $p$ on a large enough lattice. Although we can observe that the surface width has a logarithmic dependence on $t$ for short time, but for longer times, we get a non-linear behavior and, an even clearer discrepancy for all data (except $p=0.5$), which clearly shows that they fall into the KPZ universality class. Since the KPZ term, among all linear and nonlinear growth terms, is more relevant than the EW term, the deviation from EW is evidence of crossover to the KPZ regime in the hydrodynamic limit. We believe that this provides very strong evidence against the validity of the roughening transition guess for $p<0.5$.

\section{The interface velocity}
\label{Sec::interface-velocity}
In some growth models, such as the SS model, it is difficult to obtain reliable scaling exponents, due to complicated crossover and finite-size effects. An alternative method for identifying the universality class is to obtain direct evidence for the presence of different terms in the growth equation. The determination of the coefficient $\lambda$ is of special interest since, if present, $\lambda$ controls the scaling properties of the interface. The simplest method of obtaining information on the existence of the nonlinear KPZ term affecting growth processes is based on the fact that the average interface velocity, $v\equiv d \left\langle h \right\rangle/dt $, depends on both the {\it interface orientation} and {\it finite-size} \cite{Barabasi95,Krug89,Krug90}. 
A central characteristic of KPZ class is the lateral growth that results in an excess interface velocity for a substrate with an overall tilt of slope $m \equiv \left\langle \nabla h \right\rangle $. Based on this fact, the tilt method, as a powerful tool, was initially proposed by Krug \cite{Krug89,Krug90} to evaluate the nonlinearity of the associated equation for a discrete growth model. When $\abs{m}\ll 1$, there is a simple dependence between the interface velocity and slope $m$ \cite{Barabasi95}
\begin{equation}
v(m,L\rightarrow\infty)=v(0,L\rightarrow\infty)+\frac{\lambda}{2}m^2
\label{eq::lambda_calculation}
\end{equation}
\noindent where $v(0,L\rightarrow\infty)$ is interface velocity for untilted lattice in the hydrodynamic limit. The parameter $\lambda$ in SS model can be determined using deposition on tilted large substrates with an overall slope $m$. For this purpose, we can generate an overall slope $m $ of the interface by tilting the surface. Operationally, this can be performed by applying the helical boundary conditions \cite{Barabasi95}, {\it i.e.} $h(L,t) = h(1,t)-m(L-1)$. 
Based on an approach known as the Krug-Meakin method~\cite{Krug90}, it is expected for the KPZ equation that the asymptotic velocity $v_L$ for finite systems of size L is given by \cite{Krug90_Math}
\begin{equation}
\Delta v=v_L-v_{\infty}=-\frac{A\lambda}{2}L^{2\alpha-2}
\label{eq::asymp_velocity}
\end{equation}
\noindent where $A\sim D/\nu$ is the power-law coefficient of the second-order height-difference correlation as a function of the distance between columns.
In the following, after a general description of the methods, we try to estimate the interface velocity as well as the nonlinear parameter associated with the KPZ equation for the SS model. For this purpose, we begin with the determination of the interface velocity. In the $d-$dimensional substrate, we consider $\mathcal{P}^{+}$ ($\mathcal{P}^{-}$) as the probability of choosing a site eligible for growth (desorption). Since the interface height for each allowed growth (desorption) site increases (decreases) by 2, the interface velocity is given by the relation~\cite{Barabasi95}
\begin{equation}
v(t) = 2 \left[p \mathcal{P}^{+}(t) -q\mathcal{P}^{-}(t)\right]
\label{eq::velocity_ssm}
\end{equation}

\begin{figure}[!]
\begin{center}
\includegraphics[width=0.48\textwidth]{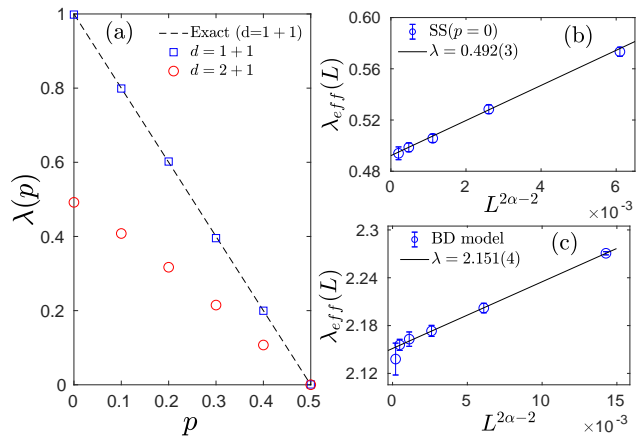}
\caption{(Color online) \textbf{(a)} The nonlinear parameter $\lambda$ of the SS model as a function of the value of $p$ in both $1+1$, and $2+1$ dimensions (the error bars are smaller than the symbols).
The dashed line is plotted based on exact theoretical results in $d=1+1$. In contrast to $d=1+1$, the obtained results in $d=2+1$ have not linear relationship with $p$. The plot of $\lambda_{eff}(L)$ against $L^{2\alpha-2}$, \textbf{(b)} for SS model at $p=0$, and \textbf{(c)} for the $(2+1)$-dimensional BD model.
}
\label{fig::lambda}
\end{center}
\end{figure}

\begin{figure}[!]
\begin{center}
\includegraphics[width=0.48\textwidth]{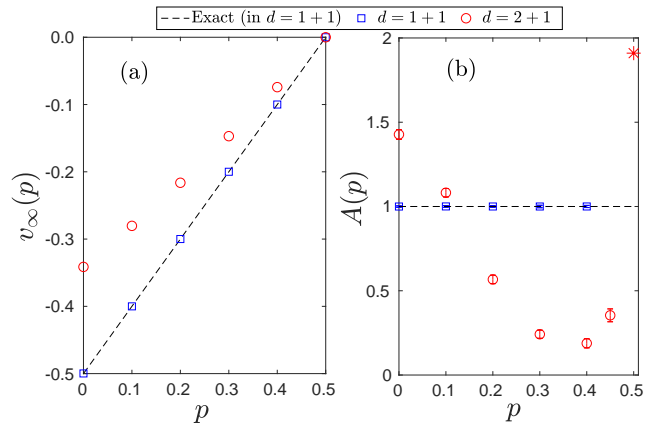}
\caption{(Color online) \textbf{(a)} The average interface velocity $v_{\infty}$, and \textbf{(b)} the $A$ parameter of the SS model vs the value of $p$ in both $1+1$, and $2+1$ dimensions. The maximum lattice sizes are $2^{13}$ and $2^{10}$ for $1+1$ and $2+1$, respectively.  The $\ast$ symbol denotes the exact theoretical value $\frac{6}{\pi}$ in $d=2+1$. The $v_{\infty}$ values and their statistical error are presented in Table \ref{tab:parameters}.
}
\label{fig::velocity}
\end{center}
\end{figure}

In $d=1+1$, there is a standard mapping between the height in the SS model and a kinetic Ising model \cite{Meakin86,Plischke87}. By using one essential property of the kinetic Ising model that in its steady-state all spin configurations are equivalent, the exact value of the probability of choosing a site eligible for growth (desorption) in the steady-state is given as~\cite{Krug92}
\begin{equation}
\mathcal{P}^{+}_{\infty}=\mathcal{P}^{-}_{\infty} =\frac{1}{4}(1+\frac{1}{L-1} )
\end{equation}
\noindent where $\mathcal{P}^{+}_{\infty}$ and $\mathcal{P}^{-}_{\infty}$ are the steady-state values of the $\mathcal{P}^{+}$, and $\mathcal{P}^{-}$, respectively. After substitution of these values into Eq.~(\ref{eq::velocity_ssm}), one obtains 
\begin{equation}
v_L =v_{\infty}+\frac{(p  -q)}{2}\frac{1}{L}
\label{eq::ss_velocity}
\end{equation}
\noindent where $v_{\infty}=\frac{1}{2}(p-q)$ is the asymptotic velocity of the interface.
On the other hand, by tilting the substrate, the exact analytic result for the coefficient of the nonlinear term in the KPZ equation is known as $\lambda =(q-p)$ \cite{Barabasi95,Krug92}. This relation expresses quantitatively the fact that only for $p=q$, the nonlinear term vanishes, and the SS model belongs to the EW class which is in excellent agreement with our previous numerical observations in interface width. Comparing Eq.~(\ref{eq::ss_velocity}), and Eq.~(\ref{eq::asymp_velocity}) with $\lambda =(q-p)$ conclude to $A=1$, independent of the value of $p$.
It should be noted that the exact values of $A$ and $\Gamma$ for the SS model at $p=0$ are given in Ref.~\cite{Krug92}, in this paper, we simply calculate these parameters for other values of $p$.
In Fig.~\ref{fig::lambda}, and Fig.~\ref{fig::velocity}, exact theoretical values (dashed line) and our numerical results (squares) are presented. There is an excellent agreement between the theoretical and numerical results for all values of $p$.

In $d=2+1$, in contrast to the exact results in $d=1+1$, the scenario is more complicated, although it is known that the SS interface can be mapped onto the six-vertex model with equal vertex energies \cite{Meakin86, Gwa92}, but, to our knowledge, this map has not provided any precise result about the universal and nonuniversal parameters of this model, yet.
Therefore, we try to numerically obtain the probability of finding a site eligible for growth (deposition), {\it i.e.} $\mathcal{P}^{+}_{\infty}$ ($\mathcal{P}^{-}_{\infty}$), in the steady-state regime ($t\gg L^{z}$) on a lattice of size $1024$. In Table \ref{tab:parameters}, we display the obtained values together with their statistical error of the $\mathcal{P}^{+}_{\infty}$ and $\mathcal{P}^{-}_{\infty}$. As can be seen, these probabilities are numerically equal to each other only for the case $p=q$, which, based on some symmetry principles, the model must be described by the EW equation. This finding is likely to be inconsistent with the claim that all possible configurations of the six-vertex model equally are weighted.
We believe that the $\mathcal{P}^{+}_{\infty}$ and the $\mathcal{P}^{-}_{\infty}$ are obviously related to the number of maxima and minima on the interface, as features of the local geometry, and consequently are related to the height distributions (HDs) of the surface.

A matter of concern, when obtaining numerically the $\lambda$ parameter for the SS model as well as other growth models, is related to the lattice size, because, as mentioned before, we must perform our numerical simulations on a large lattice size.
To reduce the finite-size effects in our numerical results, based on Eq.~(\ref{eq::lambda_calculation}) and Eq.~(\ref{eq::asymp_velocity}), we can estimate the asymptotic interface velocity for tilted substrates, {\it i.e.} $m\neq 0$, and then we can obtain the $\lambda$ parameter for the SS model. Consequently, for a lattice of size $L$, we can obtain the following relation for the effective nonlinear parameter:
\begin{equation}
\lambda_{eff}(L)=\lambda-B L^{2\alpha-2}
\label{eq::lambda_eff}
\end{equation}
\noindent where $\lambda$ and $B$, respectively, are the nonlinear parameter of the associated KPZ equation in the thermodynamic limit, and a constant related to the $A$ parameter.
We measure the interface velocity $v(m,L)$, as described in \cite{Krug89,Krug90}, using deposition on a tilted substrate of size $L$ with an overall slope $m$ (where $m< 0.25$). Then, based on Eq.~\ref{eq::lambda_calculation}, by fitting a parabola to the obtained interface velocities, we obtain $\lambda_{eff}(L)$ for each lattice size.
By plotting $\lambda_{eff}(L)$ against $L^{2\alpha -2}$ with the value $\alpha=0.3889(3)$ which is adopted as the roughness exponent for the KPZ class in $d=2+1$~\cite{Kelling17}, we determine $\lambda$ as listed in Table \ref{tab:parameters}. In contrast to $d=1+1$, the obtained results in $d=2+1$ do not have a linear relationship with $p$ (as shown in Fig.~\ref{fig::lambda}(a) ).
It is worth to mention that the same type of behavior in Fig.~\ref{fig::lambda}(a) has been reported in some $(1+1)$-dimensional competitive models \cite{Chame02,Muraca04,Silveira12}.
To demonstrate the accuracy and efficiency of Eq.~(\ref{eq::lambda_eff}), we also perform simulations on the BD model and estimate the nonlinear parameter of this model (as shown in Fig.~\ref{fig::lambda}(c) ).   
In a small amount of computational time, we obtain $\lambda=1.283(2)$, and $2.151(4)$ in $1+1$, and $2+1$ dimensions, respectively, which are in unprecedented accuracy compared to reported values of $1.25$ \cite{Torres17}, $1.30$ \cite{Krug92}, and $1.34$ \cite{Oliveira12PRE} for $d=1+1$, and $2.15(10)$\cite{Alves14} for $d=2+1$.

By using Eq.~(\ref{eq::ss_velocity}) and the obtained probabilities of $\mathcal{P}^{+}_{\infty}$ and $\mathcal{P}^{-}_{\infty}$, we can directly calculate the interface velocity, but to reduce the finite-size effects, we apply the Eq.~(\ref{eq::asymp_velocity}) in our numerical simulations. Therefore, by plotting $v_L$ against $L^{2\alpha -2}$, and by using the $\lambda$ parameters, we determine $v_{\infty}$ (as listed in Table \ref{tab:parameters}), and $A$.
Fig.~\ref{fig::velocity} shows a nonlinear dependence on the parameter $p$ for both $v_{\infty}$ and $A$ in $2+1$ dimensions. However, since $v_{\infty}$ vanishes at $p=0.5$, the $A$ parameter could not be determined numerically, but, fortunately, the exact value of this parameter, as discussed in Sec.~\ref{Sec::surface_width}, is $\frac{6}{\pi}$. To confirm this prediction for $p=0.5$, as shown in Fig.~\ref{fig::velocity}(b), we also calculate the $A$ parameter at $p=0.45$. This parameter exhibits a decreasing trend up to $p\approx 0.4$ and then increases toward $\frac{6}{\pi}$.

Although, for large values of $p$, as shown in Fig.~\ref{fig::lambda}(a) and Fig.~\ref{fig::velocity}(a), both of $\lambda$ and $v_{\infty}$ behave linearly in $p'$, there are deviations from the linear behavior if $p$ is not large (for instance, between $0$ and $0.25$).
To estimate the nonlinear dependence on $p$, we determine the $\lambda$ parameter and the average interface velocity $v_{\infty}$ for each of several probabilities ($p=0$, $0.05$, $...$, $0.25$). Then, we perform the least-square regression fits of the forms $\lambda\sim p'^{\gamma}$ and $v_{\infty}\sim p'^{\delta}$. Considering the obtained results (not shown in the figures) and their statistical error bars, we obtain the exponents $\gamma=0.87(4)$ and $\delta=0.90(2)$.
 
\section{Universal and non-universal parameters}
\label{Sec::non-universal}

\begin{table}[!] 
\begin{tabular}{cc|cccccccccc}\hline\hline
$d$&$[p]$  & $\mathcal{P}^{+}_{\infty}$  & $\mathcal{P}^{-}_{\infty}$ & $v_\infty$  & $\lambda$ &$\Gamma$ \\\hline
$1+1$&$[p]$  & $\frac{1}{4}$ & $\frac{1}{4}$ &  $\frac{1}{2}(p-q)$  & $(q-p)$   & $\frac{1}{2}(q-p)$\\ \hline 
$2+1$&$[0.0]$& 0.19756(4) & 0.17068(6)  & -0.34137(7)  & 0.492(3) & 1.23(6) \\
$2+1$&$[0.1]$& 0.20237(3) & 0.17825(3)  & -0.2804(1)  & 0.410(2) & 0.50(3) \\
$2+1$&$[0.2]$& 0.20790(3) & 0.18719(4)  & -0.2163(2)  & 0.317(1) & 0.074(8) \\
$2+1$&$[0.3]$& 0.21131(4) & 0.19562(5)  & -0.1471(3)  & 0.216(2) & 0.005(1) \\
$2+1$&$[0.4]$& 0.21104(6) & 0.20253(4)  & -0.0742(2)  & 0.108(1) & 0.0015(5) \\
$2+1$&$[0.5]$& 0.20789(3) & 0.20790(4)  & $\approx 0$  &  $\approx 0$ & $\approx 0$ \\\hline\hline
\end{tabular}
\caption{ Non-universal parameters for the SS model in both $1+1$ and $2+1$ dimensions at different $p$ values which are shown in brackets. In the case of $p=0$, ignoring the sign, the obtained $\nu_{\infty}$ value is in good agreement with $0.341368(3)$ reported in \cite{Oliveira13}.}
\label{tab:parameters}
\end{table}

\begin{figure}[!t]
\begin{center}
\includegraphics[width=0.48\textwidth]{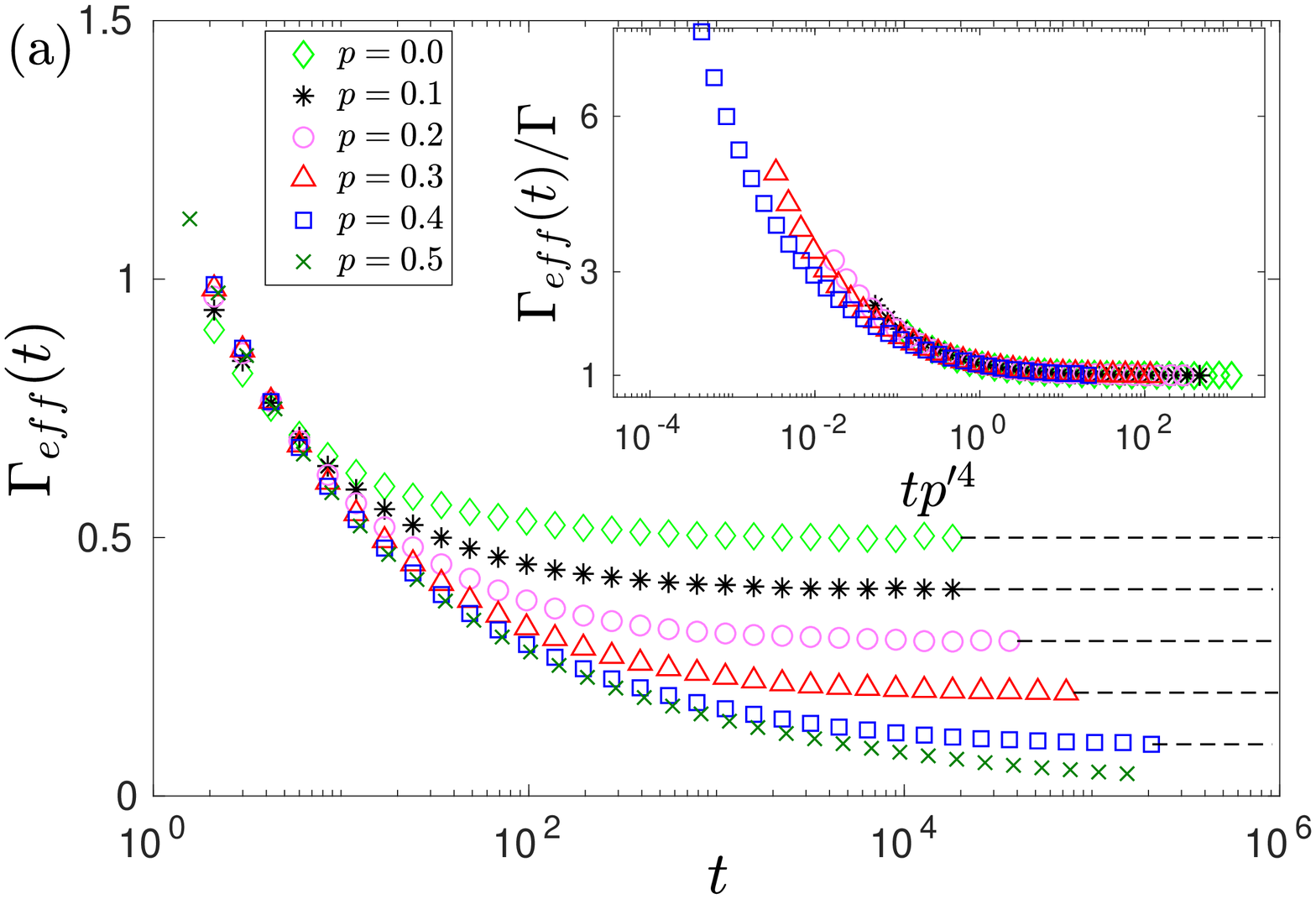}
\includegraphics[width=0.48\textwidth]{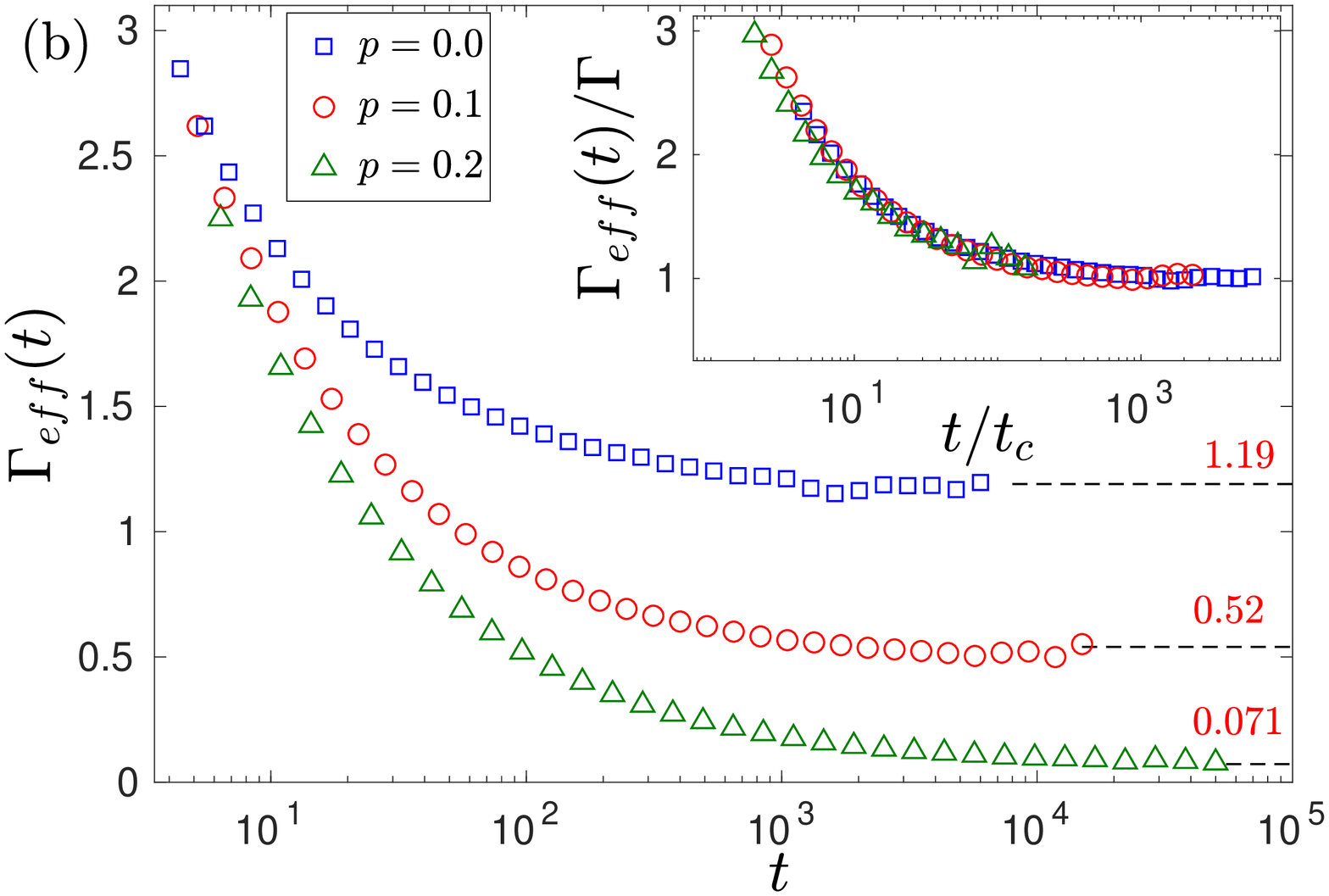}
\caption{(Color online) Amplitude fluctuation parameter estimated via the KPZ {\it ansatz} for the SS model in both $1+1$ \textbf{(a)}, and $2+1$ \textbf{(b)} dimensions. The lattice size is $2^{15}$, and $2^{10}$ for $d=1+1$, and $d=2+1$ respectively. The dashed horizontal lines are at $\Gamma$ values given by exact value ($d=1+1$) and extrapolation of $\Gamma_{eff}$ in the large time limit in $d=2+1$. The insets show rescaled $\Gamma_{eff}(t)/\Gamma$ vs $t/t_c$.} 
\label{fig::gamma}
\end{center}
\end{figure}

The scaling analysis based on the KPZ {\it ansatz}, Eq.~(\ref{eq::ansatz}), requires precise estimates of both the universal and the nonuniversal parameters. In this section, we first estimate the non-universal parameter $\Gamma$ in Eq.~(\ref{eq::ansatz}) which is controlling the amplitude of fluctuations in the KPZ {\it ansatz}. Then we investigate the universal properties of $\chi$ in both the growth and the stationary regimes.
According to an approach which is commonly called Krug-Meakin method~\cite{Krug90}, and based on the definitions adopted in past studies, the parameter $\Gamma$ is given by $\Gamma =(1/2)|\lambda |A^2$ for $1+1$ dimensions and $\Gamma =|\lambda |{{A}^{1/\alpha }}$ for $2+1$ dimensions~\cite{Krug92,Krug90}. The parameter $A$ can be obtained from the foregoing expression of the asymptotic velocity $v_L$, {\it i.e.} Eq.~(\ref{eq::asymp_velocity}).
In $d=1+1$, accepting $A =1$, and $\lambda =(q-p)$ result in $\Gamma=(q-p)/2$. For $(2+1)$-dimensional SS model, we numerically determine the parameter of $\Gamma$ for different values of $p$. These estimated values of $\Gamma$ for different values of $p$ are shown in Table~\ref{tab:parameters}. 
Moreover, to obtain $\Gamma$, there is another method which is directly related to the KPZ {\it ansatz} in the growth regime, Eq.~(\ref{eq::ansatz}), $\Gamma$ can also be obtained using
\begin{equation}
\Gamma=\lim_{t\rightarrow\infty} 
\left[\frac{\lrangle{h^2}_c}{t^{2\beta}\lrangle{\chi^2}_c}\right]^{1/2\beta},
\label{eq::Gamma}
\end{equation}
\noindent where we use $\lrangle{\chi^2}_c =0.63805$ in $1+1$ dimensions~\cite{Prahofer00}, and $\lrangle{\chi^2}_c=0.235$ in $2+1$ dimensions~\cite{Halpin12,Halpin13}. We also adopt $\beta=0.2414(2)$ as the KPZ growth exponent in $d=2+1$, which is adopted as the growth exponent based on the recent numerical estimation for the KPZ class in $d=2+1$ \cite{Kelling17}.
To consider the finite-time effects on $\Gamma$, from Eq.~(\ref{eq::ansatz}) we define
\begin{equation}
\Gamma_{eff}(t) \equiv \left[\frac{\lrangle{h^2}_c}{t^{2\beta}\lrangle{\chi^2}_c}\right]^{1/2\beta} 
= \Gamma+ct^{-2\beta}+\cdots.
\label{eq::Gamma2}
\end{equation}

\begin{figure}[!t]
\begin{center}
\includegraphics[width=0.48\textwidth]{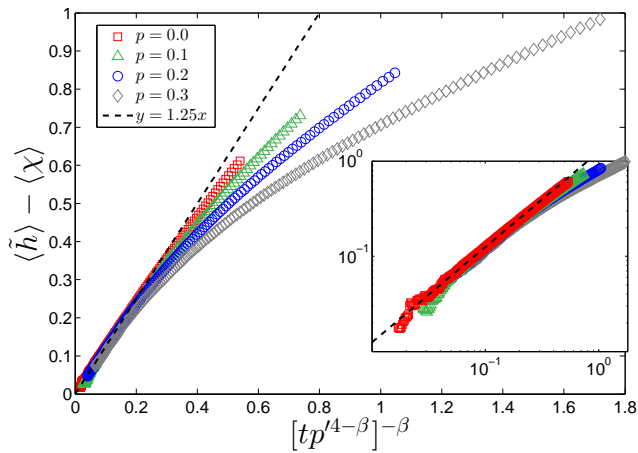}
\caption{(Color online) The variation of the $\lrangle{\tilde{h}}-\lrangle{\chi}$ vs rescaled time $\left[ tp^{\prime 4-\beta}\right]^{-\beta}$ for a lattice size  $2^{15}$ for different values of $p$ in $(1+1)$-dimensional SS model. The inset shows the same data in a log-log plot.
}
\label{fig::eta}
\end{center}
\end{figure}

The Fig.~\ref{fig::gamma} shows $\Gamma_{eff}(t)$ as a function of time for the SS model, in both $1+1$ and $2+1$ dimensions, on a lattice of size $2^{15}$ in $d=1+1$ and $2^{10}$ in $d=2+1$.
For the large value of $p$, the linear regime expected in the KPZ {\it ansatz} is observed only for a very long time. As shown in the insets of Fig.~\ref{fig::gamma}, after the crossover time scale, the KPZ clearly dominates in the growing regime with the predicted distributions.
The asymptotic $\Gamma$ values obtained using this approach are the same, inside the error bars, as those found using the Krug-Meakin analysis shown in Table~\ref{tab:parameters}. 
Although so far, we have estimated all the parameters of Eq.~(\ref{eq::ansatz}) which is valid in the limit of $t\rightarrow\infty$, but in the finite-time scale, some other nonuniversal parameters are also required to be added to that equation.
In particular, it has been reported that the first cumulant of the scaled height $\tilde{h}\equiv (h-v_{\infty}t)/(s_{\lambda}(\Gamma t)^{\beta})$ approaches the theoretical value of associated distributions as a power-law $t^{-\beta}$, {\it i.e.} $\lrangle{\tilde{h}}-\lrangle{\chi}\sim t^{-\beta}$ (for example see \cite{Amir11,Ferrari11,Takeuchi12,Calabrese11,Oliveira13,Alves14}). By adding a model-dependent stochastic quantity, such as $\eta$ responsible for a shift in the mean of the scaled height $\tilde{h}$, to the Eq.~(\ref{eq::ansatz}), a modified KPZ {\it ansatz} in the finite-time regime can be obtained. Interestingly, one can obtain the exact analytical form of KPZ {\it ansatz}, Eq~(\ref{eq::ansatz}), for $(1+1)$-dimensional SS model in the KPZ-regime:
\begin{equation}
h(t) \simeq \frac{(p-q)}{2}t + s_{\lambda} \left( \frac{q-p}{2} \right) ^{1/3}t^{1/3} \chi +\eta, 
\label{eq::1d_ssm_ansatz}
\end{equation}
where $s_{\lambda}=Sgn(q-p)$ is the sign of the $\lambda$. This exact analytical expression can be used to verify different numerical algorithms. The mean $\lrangle{\eta}$ can be determined using the
height scaled in terms of exact values of the parameters $v_\infty$ and $\Gamma$ as $\lrangle{\tilde{h}}-\lrangle{\chi}=\frac{\lrangle{\eta}}{s_{\lambda}(\Gamma)^{\beta}}t^{-\beta}$. Here we use $\lrangle{\chi} =-0.76007$ in $1+1$ dimensions~\cite{Prahofer00}. Fig.~\ref{fig::eta} shows that the power-law $t^{-\beta}$ describes very precisely the shift. So, using the prefactor of the power-law $t^{-\beta}$, we can determine $\lrangle{\eta}$ as a function of $p$. To obtain a good data collapse, both time $t$ and parameter $\Gamma$ should scale with $p^{\prime}$ for other values of $p$ with respect to the case of $p=0$. The former and the later need to scale with $p^{\prime 4}$, and $p^{\prime \beta}$, consequently the time in the prefactor needs to scale with a factor of $p^{\prime 4-\beta}$. Therefore, by applying this appropriate scaling, we expect a good data collapse, as shown in Fig.~\ref{fig::eta}. Using the prefactor of the power law $t^{\beta}$, finally, we can estimate the mean value of $\eta$, 
\begin{equation}
\lrangle{\eta}\approx\frac{\Omega}{2^{\frac{1}{3}}}(q-p)^{\frac{-8}{9}}
\label{eq::1d_ssm_ansatz2}
\end{equation}
\noindent where $\Omega=1.243(8)\approx \frac{5}{4}$ is a constant which is theoretically unknown at the present time, but can be estimated from the slope of the fitted curve in the main panel of Fig.~\ref{fig::eta}.
It is important to mention that the KPZ {\it ansatz}, {\it i.e.} Eq.~\ref{eq::ansatz}, as well as the equations derived from it (Eq.~\ref{eq::Gamma}-Eq.~\ref{eq::1d_ssm_ansatz2}) are valid only in the growth regime.

\begin{figure}[t!]
\begin{center}
\includegraphics[width=8.1cm,height=11.0cm,keepaspectratio]{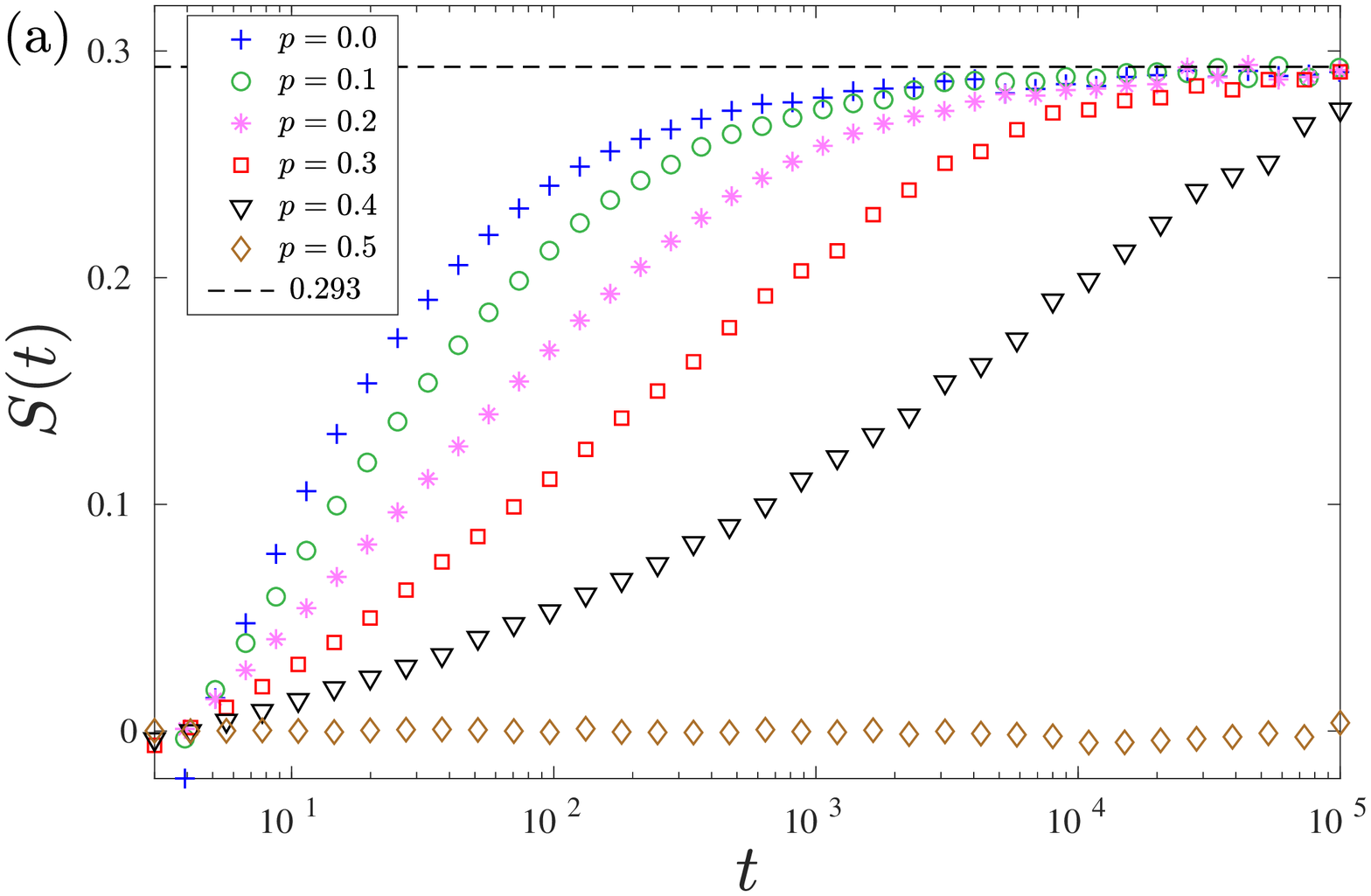}
\includegraphics[width=8.1cm,height=11.0cm,keepaspectratio]{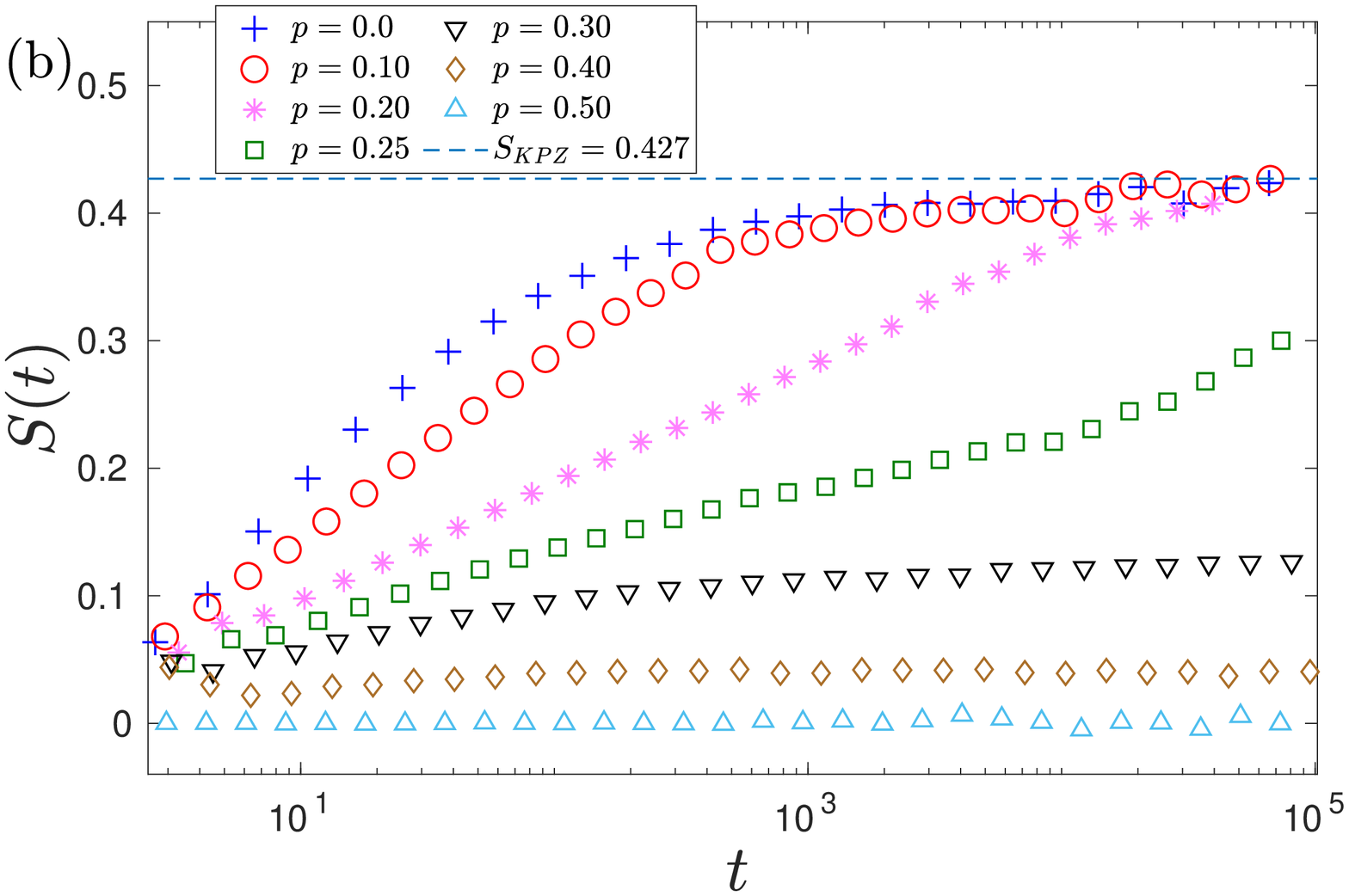}
\caption{(Color online) Skewness of HDs of the SS model for several values of $p$ in the growth regime in both $1+1$
\textbf{(a)}, and $2+1$ \textbf{(b)} dimensions. The dashed lines indicate the expected KPZ GOE values $0.2935$, and $0.427$ in $d=1+1$, and $d=2+1$ respectively. The lattice size is $2^{17}$, and $2^{11}$ for $d=1+1$, and $d=2+1$ respectively. The error bars are of the order of the size of the symbols. 
}
\label{fig::skew}
\end{center}
\end{figure}

To study the universal properties of $\chi$ of the SS model, we calculate the dimensionless cumulant ratio skewness $S=\langle h^3 \rangle_c / \langle h^2 \rangle_c^{1.5}$, which is an accurate measure of the asymmetry of the height fluctuation distribution. Here, $ \langle X^n \rangle_c$ represents the $n^{th}$ cumulant of $X$. Fig.~\ref{fig::skew} shows the skewness evolution for several values of $p$, in the growth regime in both $1+1$, and $2+1$ dimensions. Our estimated skewness $S$ for $(1+1)$-dimensional SS model, at long times, shows excellent agreement with the GOE TW distribution ($S_{\text{GOE}} = 0.2935$ \cite{Prahofer00}). In $2+1$ dimensions, as shown in Fig.~\ref{fig::skew}(b), the skewness $S$ can converge to a nonzero constant value only for the small value of $p$. The converged value \footnote{The converged value of the skewness $S_{\infty}$, in the growth regime, can be determined by performing a fit of the form $S(t)=S_{\infty}+a_1t^{-2\beta}+a_2t^{-4\beta}$ where $a_1$ and $a_2$ are free fitting parameters. We obtain $S_{\infty}=0.424(6)$ and $0.422(8)$ for $p=0$ and $0.1$, respectively, which are slightly smaller than more precise values in \cite{Oliveira13,Kelling17} due to the finite-size effects.} is in good agreement with $0.428(5)$ \cite{Oliveira13}, and $0.427(2)$ \cite{Kelling17} in the growth regime. Although due to very slow crossover for large values of $p$, which prevents the HDs from reaching the asymptotic distribution, we are not able to observe the predicted distribution in a reasonable amount of computational time, one can see that $S$ increases almost monotonically with time. Since $S= 0$ for EW growth, a small value of $S$ ( \textit{e.g.}, $S(p=0.3)\sim  0.13$), in short times, is a signature of a smooth crossover from $S_{\text{EW}}$ to $S_{\text{KPZ}}$. Thus, this suggests that the up-down symmetry is broken for any $p<0.5$ even at short times. To investigate the HDs in the stationary regime, we also calculate the skewness $S$, in particular for $p=0.15$, in which the crossover time is accessible in our simulations. The obtained skewness value in this regime is $0.26(2)$, which is in good agreement with $0.2657(4)$ \cite{Pagnani16}, and $0.270(5)$ \cite{Kelling16}. Therefore, we show that the SS model in $2+1$ dimensions obeys the KPZ {\it ansatz} with the expected universal stochastic term $\chi$.


\section{Conclusions} 
\label{Sec::Conc}

In this paper, we study the kinetic roughening of the SS model for surface growth in $1+1$ and $2+1$ dimensions.  
The results of extensive simulations, as well as our careful finite-size scaling analysis, clearly indicate the following. First, in sharp contrast to the recent report~\cite{Dashti17}, and in agreement with \cite{Barabasi95,Imbrie88, Tang90}, we show that there exists a slow crossover from an intermediate regime dominated by the EW class to an asymptotic regime dominated by the KPZ class for any $p <0.5$. Therefore, our results rule out any roughening transition in $2+1$ dimensions. Indeed, reliable estimation of the universal parameters requires appropriate consideration of the crossover from the linear behavior of the surface fluctuations at early times to the nonlinear behavior at sufficient large times. So, the presence of long crossover time for large values of $p$ leads to failure of observation of hydrodynamic limit behaviors in numerical simulations on small lattices. Second, as shown in Figs~\ref{fig::lambda} and~\ref{fig::velocity}, the effective nonuniversal parameters of $\lambda$, $v_{\infty}$, and $\Gamma$ continuously decrease with $p$, but not in a linear fashion, Finally, the universal and the nonuiversal properties of HDs of SS model also show a good agreement with the KPZ {\it ansatz}. Therefore, in the hydrodynamic limit, one expects that the growth dynamic of the SS model is described by the KPZ equation for $p\neq0.5$. 
Our study can open a new theoretical challenge in the field and can also shed light on the controversial relationship between the SS model and some extensively studied models in equilibrium or nonequilibrium statistical mechanics, such as the six-vertex model.
We believe that Eq.~(\ref{eq::lambda_eff}), and Eq.~(\ref{eq::1d_ssm_ansatz}) should be useful in numerical studies of growth models, helping to estimate with good accuracy the non-universal parameters, and to verify the numerical recipes with an exact theoretical result, respectively.  

\textbf{\textit{Note  Added:}}{A recent work \cite{Gomes19} numerically obtained the $\lambda$ parameter for an etching model up to $6+1$ dimensions and fits the data with a function of the type $\lambda(L)=\lambda-a/L^c$, where $a$ and $c$ are free fitting parameters. The numerical values of the $c$ exponent are in good agreement with the predicted values of $2-2\alpha$ presented in Eq.(~\ref{eq::lambda_eff}).}\\

\section{Acknowledgments}
We are grateful to S. Rouhani and A. Ramezanpour for helpful comments on the manuscript. We also thank F. A. Oliveira for verifying our proposed Eq.~(\ref{eq::lambda_eff}) in his numerical results \cite{Gomes19}.

\bibliography{SSM2020}

\begin{thebibliography}{63}%
\makeatletter
\providecommand \@ifxundefined [1]{%
 \@ifx{#1\undefined}
}%
\providecommand \@ifnum [1]{%
 \ifnum #1\expandafter \@firstoftwo
 \else \expandafter \@secondoftwo
 \fi
}%
\providecommand \@ifx [1]{%
 \ifx #1\expandafter \@firstoftwo
 \else \expandafter \@secondoftwo
 \fi
}%
\providecommand \natexlab [1]{#1}%
\providecommand \enquote  [1]{``#1''}%
\providecommand \bibnamefont  [1]{#1}%
\providecommand \bibfnamefont [1]{#1}%
\providecommand \citenamefont [1]{#1}%
\providecommand \href@noop [0]{\@secondoftwo}%
\providecommand \href [0]{\begingroup \@sanitize@url \@href}%
\providecommand \@href[1]{\@@startlink{#1}\@@href}%
\providecommand \@@href[1]{\endgroup#1\@@endlink}%
\providecommand \@sanitize@url [0]{\catcode `\\12\catcode `\$12\catcode
  `\&12\catcode `\#12\catcode `\^12\catcode `\_12\catcode `\%12\relax}%
\providecommand \@@startlink[1]{}%
\providecommand \@@endlink[0]{}%
\providecommand \url  [0]{\begingroup\@sanitize@url \@url }%
\providecommand \@url [1]{\endgroup\@href {#1}{\urlprefix }}%
\providecommand \urlprefix  [0]{URL }%
\providecommand \Eprint [0]{\href }%
\providecommand \doibase [0]{http://dx.doi.org/}%
\providecommand \selectlanguage [0]{\@gobble}%
\providecommand \bibinfo  [0]{\@secondoftwo}%
\providecommand \bibfield  [0]{\@secondoftwo}%
\providecommand \translation [1]{[#1]}%
\providecommand \BibitemOpen [0]{}%
\providecommand \bibitemStop [0]{}%
\providecommand \bibitemNoStop [0]{.\EOS\space}%
\providecommand \EOS [0]{\spacefactor3000\relax}%
\providecommand \BibitemShut  [1]{\csname bibitem#1\endcsname}%
\let\auto@bib@innerbib\@empty
\bibitem [{\citenamefont {Barab{\'a}si}\ and\ \citenamefont
  {Stanley}(1995)}]{Barabasi95}%
  \BibitemOpen
  \bibfield  {author} {\bibinfo {author} {\bibfnamefont {A.-L.}\ \bibnamefont
  {Barab{\'a}si}}\ and\ \bibinfo {author} {\bibfnamefont {H.~E.}\ \bibnamefont
  {Stanley}},\ }\href@noop {} {\emph {\bibinfo {title} {Fractal Concepts in
  Surface Growth}}}\ (\bibinfo  {publisher} {Cambridge University Press,
  Cambridge},\ \bibinfo {year} {1995})\BibitemShut {NoStop}%
\bibitem [{\citenamefont {Krug}(1997)}]{Krug97}%
  \BibitemOpen
  \bibfield  {author} {\bibinfo {author} {\bibfnamefont {J.}~\bibnamefont
  {Krug}},\ }\href {\doibase 10.1080/00018739700101498} {\bibfield  {journal}
  {\bibinfo  {journal} {Adv. Phys.}\ }\textbf {\bibinfo {volume} {46}},\
  \bibinfo {pages} {139} (\bibinfo {year} {1997})}\BibitemShut {NoStop}%
\bibitem [{\citenamefont {Halpin-Healy}\ and\ \citenamefont
  {Zhang}(1995)}]{Halpin-Healy95}%
  \BibitemOpen
  \bibfield  {author} {\bibinfo {author} {\bibfnamefont {T.}~\bibnamefont
  {Halpin-Healy}}\ and\ \bibinfo {author} {\bibfnamefont {Y.-C.}\ \bibnamefont
  {Zhang}},\ }\href {\doibase https://doi.org/10.1016/0370-1573(94)00087-J}
  {\bibfield  {journal} {\bibinfo  {journal} {Phys. Rep.}\ }\textbf {\bibinfo
  {volume} {254}},\ \bibinfo {pages} {215 } (\bibinfo {year}
  {1995})}\BibitemShut {NoStop}%
\bibitem [{\citenamefont {Meakin}(1998)}]{Meakin97}%
  \BibitemOpen
  \bibfield  {author} {\bibinfo {author} {\bibfnamefont {P.}~\bibnamefont
  {Meakin}},\ }\href@noop {} {\emph {\bibinfo {title} {Fractals, Scaling and
  Growth Far from Equilibrium}}}\ (\bibinfo  {publisher} {Cambridge University
  Press, Cambridge},\ \bibinfo {year} {1998})\BibitemShut {NoStop}%
\bibitem [{\citenamefont {Edwards}\ and\ \citenamefont
  {Wilkinson}(1982)}]{Edwards82}%
  \BibitemOpen
  \bibfield  {author} {\bibinfo {author} {\bibfnamefont {S.~F.}\ \bibnamefont
  {Edwards}}\ and\ \bibinfo {author} {\bibfnamefont {D.}~\bibnamefont
  {Wilkinson}},\ }\href {\doibase 10.1098/rspa.1982.0056} {\bibfield  {journal}
  {\bibinfo  {journal} {Proc. R. Soc. Lond. A}\ }\textbf {\bibinfo {volume}
  {381}},\ \bibinfo {pages} {17} (\bibinfo {year} {1982})}\BibitemShut
  {NoStop}%
\bibitem [{\citenamefont {Kardar}\ \emph {et~al.}(1986)\citenamefont {Kardar},
  \citenamefont {Parisi},\ and\ \citenamefont {Zhang}}]{Kardar86}%
  \BibitemOpen
  \bibfield  {author} {\bibinfo {author} {\bibfnamefont {M.}~\bibnamefont
  {Kardar}}, \bibinfo {author} {\bibfnamefont {G.}~\bibnamefont {Parisi}}, \
  and\ \bibinfo {author} {\bibfnamefont {Y.-C.}\ \bibnamefont {Zhang}},\ }\href
  {\doibase 10.1103/PhysRevLett.56.889} {\bibfield  {journal} {\bibinfo
  {journal} {Phys. Rev. Lett.}\ }\textbf {\bibinfo {volume} {56}},\ \bibinfo
  {pages} {889} (\bibinfo {year} {1986})}\BibitemShut {NoStop}%
\bibitem [{\citenamefont {Vold}(1959)}]{Marjorie59}%
  \BibitemOpen
  \bibfield  {author} {\bibinfo {author} {\bibfnamefont {M.~J.}\ \bibnamefont
  {Vold}},\ }\href {\doibase https://doi.org/10.1016/0095-8522(59)90041-8}
  {\bibfield  {journal} {\bibinfo  {journal} {J. Colloid Sci.}\ }\textbf
  {\bibinfo {volume} {14}},\ \bibinfo {pages} {168 } (\bibinfo {year}
  {1959})}\BibitemShut {NoStop}%
\bibitem [{\citenamefont {Kim}\ and\ \citenamefont {Kosterlitz}(1989)}]{Kim89}%
  \BibitemOpen
  \bibfield  {author} {\bibinfo {author} {\bibfnamefont {J.~M.}\ \bibnamefont
  {Kim}}\ and\ \bibinfo {author} {\bibfnamefont {J.~M.}\ \bibnamefont
  {Kosterlitz}},\ }\href {\doibase 10.1103/PhysRevLett.62.2289} {\bibfield
  {journal} {\bibinfo  {journal} {Phys. Rev. Lett.}\ }\textbf {\bibinfo
  {volume} {62}},\ \bibinfo {pages} {2289} (\bibinfo {year}
  {1989})}\BibitemShut {NoStop}%
\bibitem [{\citenamefont {Kardar}\ and\ \citenamefont
  {Zhang}(1987)}]{Kardar87}%
  \BibitemOpen
  \bibfield  {author} {\bibinfo {author} {\bibfnamefont {M.}~\bibnamefont
  {Kardar}}\ and\ \bibinfo {author} {\bibfnamefont {Y.-C.}\ \bibnamefont
  {Zhang}},\ }\href {\doibase 10.1103/PhysRevLett.58.2087} {\bibfield
  {journal} {\bibinfo  {journal} {Phys. Rev. Lett.}\ }\textbf {\bibinfo
  {volume} {58}},\ \bibinfo {pages} {2087} (\bibinfo {year}
  {1987})}\BibitemShut {NoStop}%
\bibitem [{\citenamefont {Family}\ and\ \citenamefont
  {Vicsek}(1985)}]{Family85}%
  \BibitemOpen
  \bibfield  {author} {\bibinfo {author} {\bibfnamefont {F.}~\bibnamefont
  {Family}}\ and\ \bibinfo {author} {\bibfnamefont {T.}~\bibnamefont
  {Vicsek}},\ }\href {http://stacks.iop.org/0305-4470/18/i=2/a=005} {\bibfield
  {journal} {\bibinfo  {journal} {J. Phys. A: Math. Gen.}\ }\textbf {\bibinfo
  {volume} {18}},\ \bibinfo {pages} {L75} (\bibinfo {year} {1985})}\BibitemShut
  {NoStop}%
\bibitem [{\citenamefont {Forster}\ \emph {et~al.}(1977)\citenamefont
  {Forster}, \citenamefont {Nelson},\ and\ \citenamefont
  {Stephen}}]{Forster77}%
  \BibitemOpen
  \bibfield  {author} {\bibinfo {author} {\bibfnamefont {D.}~\bibnamefont
  {Forster}}, \bibinfo {author} {\bibfnamefont {D.~R.}\ \bibnamefont {Nelson}},
  \ and\ \bibinfo {author} {\bibfnamefont {M.~J.}\ \bibnamefont {Stephen}},\
  }\href {\doibase 10.1103/PhysRevA.16.732} {\bibfield  {journal} {\bibinfo
  {journal} {Phys. Rev. A}\ }\textbf {\bibinfo {volume} {16}},\ \bibinfo
  {pages} {732} (\bibinfo {year} {1977})}\BibitemShut {NoStop}%
\bibitem [{\citenamefont {Frey}\ and\ \citenamefont {T\"auber}(1994)}]{Frey94}%
  \BibitemOpen
  \bibfield  {author} {\bibinfo {author} {\bibfnamefont {E.}~\bibnamefont
  {Frey}}\ and\ \bibinfo {author} {\bibfnamefont {U.~C.}\ \bibnamefont
  {T\"auber}},\ }\href {\doibase 10.1103/PhysRevE.50.1024} {\bibfield
  {journal} {\bibinfo  {journal} {Phys. Rev. E}\ }\textbf {\bibinfo {volume}
  {50}},\ \bibinfo {pages} {1024} (\bibinfo {year} {1994})}\BibitemShut
  {NoStop}%
\bibitem [{\citenamefont {Schwartz}\ and\ \citenamefont
  {Edwards}(1992)}]{Moshe92}%
  \BibitemOpen
  \bibfield  {author} {\bibinfo {author} {\bibfnamefont {M.}~\bibnamefont
  {Schwartz}}\ and\ \bibinfo {author} {\bibfnamefont {S.~F.}\ \bibnamefont
  {Edwards}},\ }\href {http://stacks.iop.org/0295-5075/20/i=4/a=003} {\bibfield
   {journal} {\bibinfo  {journal} {EPL (Europhys. Lett.)}\ }\textbf {\bibinfo
  {volume} {20}},\ \bibinfo {pages} {301} (\bibinfo {year} {1992})}\BibitemShut
  {NoStop}%
\bibitem [{\citenamefont {Kloss}\ \emph {et~al.}(2012)\citenamefont {Kloss},
  \citenamefont {Canet},\ and\ \citenamefont {Wschebor}}]{Kloss12}%
  \BibitemOpen
  \bibfield  {author} {\bibinfo {author} {\bibfnamefont {T.}~\bibnamefont
  {Kloss}}, \bibinfo {author} {\bibfnamefont {L.}~\bibnamefont {Canet}}, \ and\
  \bibinfo {author} {\bibfnamefont {N.}~\bibnamefont {Wschebor}},\ }\href
  {\doibase 10.1103/PhysRevE.86.051124} {\bibfield  {journal} {\bibinfo
  {journal} {Phys. Rev. E}\ }\textbf {\bibinfo {volume} {86}},\ \bibinfo
  {pages} {051124} (\bibinfo {year} {2012})}\BibitemShut {NoStop}%
\bibitem [{\citenamefont {Pagnani}\ and\ \citenamefont
  {Parisi}(2015)}]{Pagnani16}%
  \BibitemOpen
  \bibfield  {author} {\bibinfo {author} {\bibfnamefont {A.}~\bibnamefont
  {Pagnani}}\ and\ \bibinfo {author} {\bibfnamefont {G.}~\bibnamefont
  {Parisi}},\ }\href {\doibase 10.1103/PhysRevE.92.010101} {\bibfield
  {journal} {\bibinfo  {journal} {Phys. Rev. E}\ }\textbf {\bibinfo {volume}
  {92}},\ \bibinfo {pages} {010101(R)} (\bibinfo {year} {2015})}\BibitemShut
  {NoStop}%
\bibitem [{\citenamefont {Kelling}\ \emph {et~al.}(2016)\citenamefont
  {Kelling}, \citenamefont {\'Odor},\ and\ \citenamefont
  {Gemming}}]{Kelling16}%
  \BibitemOpen
  \bibfield  {author} {\bibinfo {author} {\bibfnamefont {J.}~\bibnamefont
  {Kelling}}, \bibinfo {author} {\bibfnamefont {G.}~\bibnamefont {\'Odor}}, \
  and\ \bibinfo {author} {\bibfnamefont {S.}~\bibnamefont {Gemming}},\ }\href
  {\doibase 10.1103/PhysRevE.94.022107} {\bibfield  {journal} {\bibinfo
  {journal} {Phys. Rev. E}\ }\textbf {\bibinfo {volume} {94}},\ \bibinfo
  {pages} {022107} (\bibinfo {year} {2016})}\BibitemShut {NoStop}%
\bibitem [{\citenamefont {Kelling}\ \emph {et~al.}(2018)\citenamefont
  {Kelling}, \citenamefont {\'Odor},\ and\ \citenamefont
  {Gemming}}]{Kelling17}%
  \BibitemOpen
  \bibfield  {author} {\bibinfo {author} {\bibfnamefont {J.}~\bibnamefont
  {Kelling}}, \bibinfo {author} {\bibfnamefont {G.}~\bibnamefont {\'Odor}}, \
  and\ \bibinfo {author} {\bibfnamefont {S.}~\bibnamefont {Gemming}},\ }\href
  {http://stacks.iop.org/1751-8121/51/i=3/a=035003} {\bibfield  {journal}
  {\bibinfo  {journal} {J. Phys. A: Math. Theor.}\ }\textbf {\bibinfo {volume}
  {51}},\ \bibinfo {pages} {035003} (\bibinfo {year} {2018})}\BibitemShut
  {NoStop}%
\bibitem [{\citenamefont {Johansson}(2000)}]{Johansson00}%
  \BibitemOpen
  \bibfield  {author} {\bibinfo {author} {\bibfnamefont {K.}~\bibnamefont
  {Johansson}},\ }\href {\doibase https://doi.org/10.1007/s002200050027}
  {\bibfield  {journal} {\bibinfo  {journal} {Comm. Math. Phys.}\ }\textbf
  {\bibinfo {volume} {209}},\ \bibinfo {pages} {437} (\bibinfo {year}
  {2000})}\BibitemShut {NoStop}%
\bibitem [{\citenamefont {Meakin}\ \emph {et~al.}(1986)\citenamefont {Meakin},
  \citenamefont {Ramanlal}, \citenamefont {Sander},\ and\ \citenamefont
  {Ball}}]{Meakin86}%
  \BibitemOpen
  \bibfield  {author} {\bibinfo {author} {\bibfnamefont {P.}~\bibnamefont
  {Meakin}}, \bibinfo {author} {\bibfnamefont {P.}~\bibnamefont {Ramanlal}},
  \bibinfo {author} {\bibfnamefont {L.~M.}\ \bibnamefont {Sander}}, \ and\
  \bibinfo {author} {\bibfnamefont {R.~C.}\ \bibnamefont {Ball}},\ }\href
  {\doibase 10.1103/PhysRevA.34.5091} {\bibfield  {journal} {\bibinfo
  {journal} {Phys. Rev. A}\ }\textbf {\bibinfo {volume} {34}},\ \bibinfo
  {pages} {5091} (\bibinfo {year} {1986})}\BibitemShut {NoStop}%
\bibitem [{\citenamefont {Plischke}\ \emph {et~al.}(1987)\citenamefont
  {Plischke}, \citenamefont {R\'acz},\ and\ \citenamefont {Liu}}]{Plischke87}%
  \BibitemOpen
  \bibfield  {author} {\bibinfo {author} {\bibfnamefont {M.}~\bibnamefont
  {Plischke}}, \bibinfo {author} {\bibfnamefont {Z.}~\bibnamefont {R\'acz}}, \
  and\ \bibinfo {author} {\bibfnamefont {D.}~\bibnamefont {Liu}},\ }\href
  {\doibase 10.1103/PhysRevB.35.3485} {\bibfield  {journal} {\bibinfo
  {journal} {Phys. Rev. B}\ }\textbf {\bibinfo {volume} {35}},\ \bibinfo
  {pages} {3485} (\bibinfo {year} {1987})}\BibitemShut {NoStop}%
\bibitem [{\citenamefont {Liu}\ and\ \citenamefont {Plischke}(1988)}]{Liu88}%
  \BibitemOpen
  \bibfield  {author} {\bibinfo {author} {\bibfnamefont {D.}~\bibnamefont
  {Liu}}\ and\ \bibinfo {author} {\bibfnamefont {M.}~\bibnamefont {Plischke}},\
  }\href {\doibase 10.1103/PhysRevB.38.4781} {\bibfield  {journal} {\bibinfo
  {journal} {Phys. Rev. B}\ }\textbf {\bibinfo {volume} {38}},\ \bibinfo
  {pages} {4781} (\bibinfo {year} {1988})}\BibitemShut {NoStop}%
\bibitem [{\citenamefont {Kondev}\ \emph {et~al.}(2000)\citenamefont {Kondev},
  \citenamefont {Henley},\ and\ \citenamefont {Salinas}}]{Kondev00}%
  \BibitemOpen
  \bibfield  {author} {\bibinfo {author} {\bibfnamefont {J.}~\bibnamefont
  {Kondev}}, \bibinfo {author} {\bibfnamefont {C.~L.}\ \bibnamefont {Henley}},
  \ and\ \bibinfo {author} {\bibfnamefont {D.~G.}\ \bibnamefont {Salinas}},\
  }\href {\doibase 10.1103/PhysRevE.61.104} {\bibfield  {journal} {\bibinfo
  {journal} {Phys. Rev. E}\ }\textbf {\bibinfo {volume} {61}},\ \bibinfo
  {pages} {104} (\bibinfo {year} {2000})}\BibitemShut {NoStop}%
\bibitem [{\citenamefont {Tracy}\ and\ \citenamefont {Widom}(1994)}]{Tracy94}%
  \BibitemOpen
  \bibfield  {author} {\bibinfo {author} {\bibfnamefont {C.~A.}\ \bibnamefont
  {Tracy}}\ and\ \bibinfo {author} {\bibfnamefont {H.}~\bibnamefont {Widom}},\
  }\href@noop {} {\bibfield  {journal} {\bibinfo  {journal} {Commun. Math.
  Phys.}\ }\textbf {\bibinfo {volume} {159}},\ \bibinfo {pages} {151} (\bibinfo
  {year} {1994})}\BibitemShut {NoStop}%
\bibitem [{\citenamefont {Forrester}(1993)}]{Forrester93}%
  \BibitemOpen
  \bibfield  {author} {\bibinfo {author} {\bibfnamefont {P.}~\bibnamefont
  {Forrester}},\ }\href {\doibase https://doi.org/10.1016/0550-3213(93)90126-A}
  {\bibfield  {journal} {\bibinfo  {journal} {Nucl. Phys. B}\ }\textbf
  {\bibinfo {volume} {402}},\ \bibinfo {pages} {709 } (\bibinfo {year}
  {1993})}\BibitemShut {NoStop}%
\bibitem [{\citenamefont {Krug}\ \emph {et~al.}(1992)\citenamefont {Krug},
  \citenamefont {Meakin},\ and\ \citenamefont {Halpin-Healy}}]{Krug92}%
  \BibitemOpen
  \bibfield  {author} {\bibinfo {author} {\bibfnamefont {J.}~\bibnamefont
  {Krug}}, \bibinfo {author} {\bibfnamefont {P.}~\bibnamefont {Meakin}}, \ and\
  \bibinfo {author} {\bibfnamefont {T.}~\bibnamefont {Halpin-Healy}},\ }\href
  {\doibase 10.1103/PhysRevA.45.638} {\bibfield  {journal} {\bibinfo  {journal}
  {Phys. Rev. A}\ }\textbf {\bibinfo {volume} {45}},\ \bibinfo {pages} {638}
  (\bibinfo {year} {1992})}\BibitemShut {NoStop}%
\bibitem [{\citenamefont {Pr\"ahofer}\ and\ \citenamefont
  {Spohn}(2000)}]{Prahofer00}%
  \BibitemOpen
  \bibfield  {author} {\bibinfo {author} {\bibfnamefont {M.}~\bibnamefont
  {Pr\"ahofer}}\ and\ \bibinfo {author} {\bibfnamefont {H.}~\bibnamefont
  {Spohn}},\ }\href {\doibase 10.1103/PhysRevLett.84.4882} {\bibfield
  {journal} {\bibinfo  {journal} {Phys. Rev. Lett.}\ }\textbf {\bibinfo
  {volume} {84}},\ \bibinfo {pages} {4882} (\bibinfo {year}
  {2000})}\BibitemShut {NoStop}%
\bibitem [{\citenamefont {Calabrese}\ \emph {et~al.}(2010)\citenamefont
  {Calabrese}, \citenamefont {Doussal},\ and\ \citenamefont
  {Rosso}}]{Calabrese10}%
  \BibitemOpen
  \bibfield  {author} {\bibinfo {author} {\bibfnamefont {P.}~\bibnamefont
  {Calabrese}}, \bibinfo {author} {\bibfnamefont {P.~L.}\ \bibnamefont
  {Doussal}}, \ and\ \bibinfo {author} {\bibfnamefont {A.}~\bibnamefont
  {Rosso}},\ }\href {http://stacks.iop.org/0295-5075/90/i=2/a=20002} {\bibfield
   {journal} {\bibinfo  {journal} {EPL (Europhysics Letters)}\ }\textbf
  {\bibinfo {volume} {90}},\ \bibinfo {pages} {20002} (\bibinfo {year}
  {2010})}\BibitemShut {NoStop}%
\bibitem [{\citenamefont {Dotsenko}(2010)}]{Dotsenko10}%
  \BibitemOpen
  \bibfield  {author} {\bibinfo {author} {\bibfnamefont {V.}~\bibnamefont
  {Dotsenko}},\ }\href {http://stacks.iop.org/0295-5075/90/i=2/a=20003}
  {\bibfield  {journal} {\bibinfo  {journal} {EPL (Europhysics. Letters.)}\
  }\textbf {\bibinfo {volume} {90}},\ \bibinfo {pages} {20003} (\bibinfo {year}
  {2010})}\BibitemShut {NoStop}%
\bibitem [{\citenamefont {Sasamoto}\ and\ \citenamefont
  {Spohn}(2010)}]{Sasamoto10}%
  \BibitemOpen
  \bibfield  {author} {\bibinfo {author} {\bibfnamefont {T.}~\bibnamefont
  {Sasamoto}}\ and\ \bibinfo {author} {\bibfnamefont {H.}~\bibnamefont
  {Spohn}},\ }\href {\doibase 10.1103/PhysRevLett.104.230602} {\bibfield
  {journal} {\bibinfo  {journal} {Phys. Rev. Lett.}\ }\textbf {\bibinfo
  {volume} {104}},\ \bibinfo {pages} {230602} (\bibinfo {year}
  {2010})}\BibitemShut {NoStop}%
\bibitem [{\citenamefont {Amir}\ \emph {et~al.}(2011)\citenamefont {Amir},
  \citenamefont {Corwin},\ and\ \citenamefont {Quastel}}]{Amir11}%
  \BibitemOpen
  \bibfield  {author} {\bibinfo {author} {\bibfnamefont {G.}~\bibnamefont
  {Amir}}, \bibinfo {author} {\bibfnamefont {I.}~\bibnamefont {Corwin}}, \ and\
  \bibinfo {author} {\bibfnamefont {J.}~\bibnamefont {Quastel}},\ }\href
  {\doibase 10.1002/cpa.20347} {\bibfield  {journal} {\bibinfo  {journal}
  {Commun. Pure Appl. Math.}\ }\textbf {\bibinfo {volume} {64}},\ \bibinfo
  {pages} {466} (\bibinfo {year} {2011})}\BibitemShut {NoStop}%
\bibitem [{\citenamefont {Takeuchi}\ and\ \citenamefont
  {Sano}(2010)}]{Takeuchi10}%
  \BibitemOpen
  \bibfield  {author} {\bibinfo {author} {\bibfnamefont {K.~A.}\ \bibnamefont
  {Takeuchi}}\ and\ \bibinfo {author} {\bibfnamefont {M.}~\bibnamefont
  {Sano}},\ }\href {\doibase 10.1103/PhysRevLett.104.230601} {\bibfield
  {journal} {\bibinfo  {journal} {Phys. Rev. Lett.}\ }\textbf {\bibinfo
  {volume} {104}},\ \bibinfo {pages} {230601} (\bibinfo {year}
  {2010})}\BibitemShut {NoStop}%
\bibitem [{\citenamefont {Takeuchi}\ \emph {et~al.}(2011)\citenamefont
  {Takeuchi}, \citenamefont {Sano}, \citenamefont {Sasamoto},\ and\
  \citenamefont {Spohn}}]{Takeuchi11}%
  \BibitemOpen
  \bibfield  {author} {\bibinfo {author} {\bibfnamefont {K.~A.}\ \bibnamefont
  {Takeuchi}}, \bibinfo {author} {\bibfnamefont {M.}~\bibnamefont {Sano}},
  \bibinfo {author} {\bibfnamefont {T.}~\bibnamefont {Sasamoto}}, \ and\
  \bibinfo {author} {\bibfnamefont {H.}~\bibnamefont {Spohn}},\ }\href
  {http://dx.doi.org/10.1038/srep00034} {\bibfield  {journal} {\bibinfo
  {journal} {Sci. Rep. (Nature)}\ }\textbf {\bibinfo {volume} {1}},\ \bibinfo
  {pages} {34} (\bibinfo {year} {2011})}\BibitemShut {NoStop}%
\bibitem [{\citenamefont {Oliveira}\ \emph {et~al.}(2013)\citenamefont
  {Oliveira}, \citenamefont {Alves},\ and\ \citenamefont
  {Ferreira}}]{Oliveira13}%
  \BibitemOpen
  \bibfield  {author} {\bibinfo {author} {\bibfnamefont {T.~J.}\ \bibnamefont
  {Oliveira}}, \bibinfo {author} {\bibfnamefont {S.~G.}\ \bibnamefont {Alves}},
  \ and\ \bibinfo {author} {\bibfnamefont {S.~C.}\ \bibnamefont {Ferreira}},\
  }\href {\doibase 10.1103/PhysRevE.87.040102} {\bibfield  {journal} {\bibinfo
  {journal} {Phys. Rev. E}\ }\textbf {\bibinfo {volume} {87}},\ \bibinfo
  {pages} {040102(R)} (\bibinfo {year} {2013})}\BibitemShut {NoStop}%
\bibitem [{\citenamefont {Halpin-Healy}(2012)}]{Halpin12}%
  \BibitemOpen
  \bibfield  {author} {\bibinfo {author} {\bibfnamefont {T.}~\bibnamefont
  {Halpin-Healy}},\ }\href {\doibase 10.1103/PhysRevLett.109.170602} {\bibfield
   {journal} {\bibinfo  {journal} {Phys. Rev. Lett.}\ }\textbf {\bibinfo
  {volume} {109}},\ \bibinfo {pages} {170602} (\bibinfo {year}
  {2012})}\BibitemShut {NoStop}%
\bibitem [{\citenamefont {Halpin-Healy}(2013)}]{Halpin13}%
  \BibitemOpen
  \bibfield  {author} {\bibinfo {author} {\bibfnamefont {T.}~\bibnamefont
  {Halpin-Healy}},\ }\href {\doibase 10.1103/PhysRevE.88.042118} {\bibfield
  {journal} {\bibinfo  {journal} {Phys. Rev. E}\ }\textbf {\bibinfo {volume}
  {88}},\ \bibinfo {pages} {042118} (\bibinfo {year} {2013})}\BibitemShut
  {NoStop}%
\bibitem [{\citenamefont {Carrasco}\ \emph {et~al.}(2014)\citenamefont
  {Carrasco}, \citenamefont {Takeuchi}, \citenamefont {Ferreira},\ and\
  \citenamefont {Oliveira}}]{Carrasco14}%
  \BibitemOpen
  \bibfield  {author} {\bibinfo {author} {\bibfnamefont {I.~S.~S.}\
  \bibnamefont {Carrasco}}, \bibinfo {author} {\bibfnamefont {K.~A.}\
  \bibnamefont {Takeuchi}}, \bibinfo {author} {\bibfnamefont {S.~C.}\
  \bibnamefont {Ferreira}}, \ and\ \bibinfo {author} {\bibfnamefont {T.~J.}\
  \bibnamefont {Oliveira}},\ }\href {\doibase 10.1088/1367-2630/16/12/123057}
  {\bibfield  {journal} {\bibinfo  {journal} {New J. Phys.}\ }\textbf {\bibinfo
  {volume} {16}},\ \bibinfo {pages} {123057} (\bibinfo {year}
  {2014})}\BibitemShut {NoStop}%
\bibitem [{\citenamefont {Derrida}(1998)}]{Derrida98}%
  \BibitemOpen
  \bibfield  {author} {\bibinfo {author} {\bibfnamefont {B.}~\bibnamefont
  {Derrida}},\ }\href {\doibase https://doi.org/10.1016/S0370-1573(98)00006-4}
  {\bibfield  {journal} {\bibinfo  {journal} {Phys. Rep.}\ }\textbf {\bibinfo
  {volume} {301}},\ \bibinfo {pages} {65 } (\bibinfo {year}
  {1998})}\BibitemShut {NoStop}%
\bibitem [{\citenamefont {Neergaard}\ and\ \citenamefont {den
  Nijs}(1997)}]{Neergaard97}%
  \BibitemOpen
  \bibfield  {author} {\bibinfo {author} {\bibfnamefont {J.}~\bibnamefont
  {Neergaard}}\ and\ \bibinfo {author} {\bibfnamefont {M.}~\bibnamefont {den
  Nijs}},\ }\href {http://stacks.iop.org/0305-4470/30/i=6/a=019} {\bibfield
  {journal} {\bibinfo  {journal} {J. Phys. A: Math. Gen.}\ }\textbf {\bibinfo
  {volume} {30}},\ \bibinfo {pages} {1935} (\bibinfo {year}
  {1997})}\BibitemShut {NoStop}%
\bibitem [{\citenamefont {Gwa}\ and\ \citenamefont {Spohn}(1992)}]{Gwa92}%
  \BibitemOpen
  \bibfield  {author} {\bibinfo {author} {\bibfnamefont {L.-H.}\ \bibnamefont
  {Gwa}}\ and\ \bibinfo {author} {\bibfnamefont {H.}~\bibnamefont {Spohn}},\
  }\href {\doibase 10.1103/PhysRevLett.68.725} {\bibfield  {journal} {\bibinfo
  {journal} {Phys. Rev. Lett.}\ }\textbf {\bibinfo {volume} {68}},\ \bibinfo
  {pages} {725} (\bibinfo {year} {1992})}\BibitemShut {NoStop}%
\bibitem [{\citenamefont {Tang}\ \emph {et~al.}(1990)\citenamefont {Tang},
  \citenamefont {Nattermann},\ and\ \citenamefont {Forrest}}]{Tang90}%
  \BibitemOpen
  \bibfield  {author} {\bibinfo {author} {\bibfnamefont {L.-H.}\ \bibnamefont
  {Tang}}, \bibinfo {author} {\bibfnamefont {T.}~\bibnamefont {Nattermann}}, \
  and\ \bibinfo {author} {\bibfnamefont {B.~M.}\ \bibnamefont {Forrest}},\
  }\href {\doibase 10.1103/PhysRevLett.65.2422} {\bibfield  {journal} {\bibinfo
   {journal} {Phys. Rev. Lett.}\ }\textbf {\bibinfo {volume} {65}},\ \bibinfo
  {pages} {2422} (\bibinfo {year} {1990})}\BibitemShut {NoStop}%
\bibitem [{\citenamefont {Tang}\ \emph {et~al.}(1992)\citenamefont {Tang},
  \citenamefont {Forrest},\ and\ \citenamefont {Wolf}}]{Tang92}%
  \BibitemOpen
  \bibfield  {author} {\bibinfo {author} {\bibfnamefont {L.-H.}\ \bibnamefont
  {Tang}}, \bibinfo {author} {\bibfnamefont {B.~M.}\ \bibnamefont {Forrest}}, \
  and\ \bibinfo {author} {\bibfnamefont {D.~E.}\ \bibnamefont {Wolf}},\ }\href
  {\doibase 10.1103/PhysRevA.45.7162} {\bibfield  {journal} {\bibinfo
  {journal} {Phys. Rev. A}\ }\textbf {\bibinfo {volume} {45}},\ \bibinfo
  {pages} {7162} (\bibinfo {year} {1992})}\BibitemShut {NoStop}%
\bibitem [{\citenamefont {Forrest}\ and\ \citenamefont
  {Tang}(1990)}]{Forrest90}%
  \BibitemOpen
  \bibfield  {author} {\bibinfo {author} {\bibfnamefont {B.~M.}\ \bibnamefont
  {Forrest}}\ and\ \bibinfo {author} {\bibfnamefont {L.-H.}\ \bibnamefont
  {Tang}},\ }\href {\doibase 10.1103/PhysRevLett.64.1405} {\bibfield  {journal}
  {\bibinfo  {journal} {Phys. Rev. Lett.}\ }\textbf {\bibinfo {volume} {64}},\
  \bibinfo {pages} {1405} (\bibinfo {year} {1990})}\BibitemShut {NoStop}%
\bibitem [{\citenamefont {Dashti-Naserabadi}\ \emph {et~al.}(2017)\citenamefont
  {Dashti-Naserabadi}, \citenamefont {Saberi},\ and\ \citenamefont
  {Rouhani}}]{Dashti17}%
  \BibitemOpen
  \bibfield  {author} {\bibinfo {author} {\bibfnamefont {H.}~\bibnamefont
  {Dashti-Naserabadi}}, \bibinfo {author} {\bibfnamefont {A.~A.}\ \bibnamefont
  {Saberi}}, \ and\ \bibinfo {author} {\bibfnamefont {S.}~\bibnamefont
  {Rouhani}},\ }\href {http://stacks.iop.org/1367-2630/19/i=6/a=063035}
  {\bibfield  {journal} {\bibinfo  {journal} {New J. Phys.}\ }\textbf {\bibinfo
  {volume} {19}},\ \bibinfo {pages} {063035} (\bibinfo {year}
  {2017})}\BibitemShut {NoStop}%
\bibitem [{\citenamefont {Alves}\ \emph {et~al.}(2014)\citenamefont {Alves},
  \citenamefont {Oliveira},\ and\ \citenamefont {Ferreira}}]{Alves14}%
  \BibitemOpen
  \bibfield  {author} {\bibinfo {author} {\bibfnamefont {S.~G.}\ \bibnamefont
  {Alves}}, \bibinfo {author} {\bibfnamefont {T.~J.}\ \bibnamefont {Oliveira}},
  \ and\ \bibinfo {author} {\bibfnamefont {S.~C.}\ \bibnamefont {Ferreira}},\
  }\href {\doibase 10.1103/PhysRevE.90.052405} {\bibfield  {journal} {\bibinfo
  {journal} {Phys. Rev. E}\ }\textbf {\bibinfo {volume} {90}},\ \bibinfo
  {pages} {052405} (\bibinfo {year} {2014})}\BibitemShut {NoStop}%
\bibitem [{\citenamefont {Amar}\ and\ \citenamefont {Family}(1990)}]{Amar90}%
  \BibitemOpen
  \bibfield  {author} {\bibinfo {author} {\bibfnamefont {J.~G.}\ \bibnamefont
  {Amar}}\ and\ \bibinfo {author} {\bibfnamefont {F.}~\bibnamefont {Family}},\
  }\href {\doibase 10.1103/PhysRevA.41.3399} {\bibfield  {journal} {\bibinfo
  {journal} {Phys. Rev. A}\ }\textbf {\bibinfo {volume} {41}},\ \bibinfo
  {pages} {3399} (\bibinfo {year} {1990})}\BibitemShut {NoStop}%
\bibitem [{\citenamefont {Sneppen}\ \emph {et~al.}(1992)\citenamefont
  {Sneppen}, \citenamefont {Krug}, \citenamefont {Jensen}, \citenamefont
  {Jayaprakash},\ and\ \citenamefont {Bohr}}]{Sneppen92}%
  \BibitemOpen
  \bibfield  {author} {\bibinfo {author} {\bibfnamefont {K.}~\bibnamefont
  {Sneppen}}, \bibinfo {author} {\bibfnamefont {J.}~\bibnamefont {Krug}},
  \bibinfo {author} {\bibfnamefont {M.~H.}\ \bibnamefont {Jensen}}, \bibinfo
  {author} {\bibfnamefont {C.}~\bibnamefont {Jayaprakash}}, \ and\ \bibinfo
  {author} {\bibfnamefont {T.}~\bibnamefont {Bohr}},\ }\href {\doibase
  10.1103/PhysRevA.46.R7351} {\bibfield  {journal} {\bibinfo  {journal} {Phys.
  Rev. A}\ }\textbf {\bibinfo {volume} {46}},\ \bibinfo {pages} {R7351}
  (\bibinfo {year} {1992})}\BibitemShut {NoStop}%
\bibitem [{\citenamefont {Oliveira}\ \emph {et~al.}(2006)\citenamefont
  {Oliveira}, \citenamefont {Dechoum}, \citenamefont {Redinz},\ and\
  \citenamefont {Aar\~ao Reis}}]{Oliveira06}%
  \BibitemOpen
  \bibfield  {author} {\bibinfo {author} {\bibfnamefont {T.~J.}\ \bibnamefont
  {Oliveira}}, \bibinfo {author} {\bibfnamefont {K.}~\bibnamefont {Dechoum}},
  \bibinfo {author} {\bibfnamefont {J.~A.}\ \bibnamefont {Redinz}}, \ and\
  \bibinfo {author} {\bibfnamefont {F.~D.~A.}\ \bibnamefont {Aar\~ao Reis}},\
  }\href {\doibase 10.1103/PhysRevE.74.011604} {\bibfield  {journal} {\bibinfo
  {journal} {Phys. Rev. E}\ }\textbf {\bibinfo {volume} {74}},\ \bibinfo
  {pages} {011604} (\bibinfo {year} {2006})}\BibitemShut {NoStop}%
\bibitem [{\citenamefont {Nattermann}\ and\ \citenamefont
  {Tang}(1992)}]{Nattermann92}%
  \BibitemOpen
  \bibfield  {author} {\bibinfo {author} {\bibfnamefont {T.}~\bibnamefont
  {Nattermann}}\ and\ \bibinfo {author} {\bibfnamefont {L.-H.}\ \bibnamefont
  {Tang}},\ }\href {\doibase 10.1103/PhysRevA.45.7156} {\bibfield  {journal}
  {\bibinfo  {journal} {Phys. Rev. A}\ }\textbf {\bibinfo {volume} {45}},\
  \bibinfo {pages} {7156} (\bibinfo {year} {1992})}\BibitemShut {NoStop}%
\bibitem [{\citenamefont {Blote}\ and\ \citenamefont
  {Hilborst}(1982)}]{Blote82}%
  \BibitemOpen
  \bibfield  {author} {\bibinfo {author} {\bibfnamefont {H.~W.~J.}\
  \bibnamefont {Blote}}\ and\ \bibinfo {author} {\bibfnamefont {H.~J.}\
  \bibnamefont {Hilborst}},\ }\href
  {http://stacks.iop.org/0305-4470/15/i=11/a=011} {\bibfield  {journal}
  {\bibinfo  {journal} {J. Phys. A: Math. Gen.}\ }\textbf {\bibinfo {volume}
  {15}},\ \bibinfo {pages} {L631} (\bibinfo {year} {1982})}\BibitemShut
  {NoStop}%
\bibitem [{\citenamefont {\'Odor}\ \emph {et~al.}(2009)\citenamefont {\'Odor},
  \citenamefont {Liedke},\ and\ \citenamefont {Heinig}}]{Odor09}%
  \BibitemOpen
  \bibfield  {author} {\bibinfo {author} {\bibfnamefont {G.}~\bibnamefont
  {\'Odor}}, \bibinfo {author} {\bibfnamefont {B.}~\bibnamefont {Liedke}}, \
  and\ \bibinfo {author} {\bibfnamefont {K.-H.}\ \bibnamefont {Heinig}},\
  }\href {\doibase 10.1103/PhysRevE.79.021125} {\bibfield  {journal} {\bibinfo
  {journal} {Phys. Rev. E}\ }\textbf {\bibinfo {volume} {79}},\ \bibinfo
  {pages} {021125} (\bibinfo {year} {2009})}\BibitemShut {NoStop}%
\bibitem [{\citenamefont {Krug}(1989)}]{Krug89}%
  \BibitemOpen
  \bibfield  {author} {\bibinfo {author} {\bibfnamefont {J.}~\bibnamefont
  {Krug}},\ }\href {http://stacks.iop.org/0305-4470/22/i=16/a=002} {\bibfield
  {journal} {\bibinfo  {journal} {J. Phys. A: Math. Gen.}\ }\textbf {\bibinfo
  {volume} {22}},\ \bibinfo {pages} {L769} (\bibinfo {year}
  {1989})}\BibitemShut {NoStop}%
\bibitem [{\citenamefont {Krug}\ and\ \citenamefont {Spohn}(1990)}]{Krug90}%
  \BibitemOpen
  \bibfield  {author} {\bibinfo {author} {\bibfnamefont {J.}~\bibnamefont
  {Krug}}\ and\ \bibinfo {author} {\bibfnamefont {H.}~\bibnamefont {Spohn}},\
  }\href {\doibase 10.1103/PhysRevLett.64.2332} {\bibfield  {journal} {\bibinfo
   {journal} {Phys. Rev. Lett.}\ }\textbf {\bibinfo {volume} {64}},\ \bibinfo
  {pages} {2332} (\bibinfo {year} {1990})}\BibitemShut {NoStop}%
\bibitem [{\citenamefont {Krug}\ and\ \citenamefont
  {Meakin}(1990)}]{Krug90_Math}%
  \BibitemOpen
  \bibfield  {author} {\bibinfo {author} {\bibfnamefont {J.}~\bibnamefont
  {Krug}}\ and\ \bibinfo {author} {\bibfnamefont {P.}~\bibnamefont {Meakin}},\
  }\href {http://stacks.iop.org/0305-4470/23/i=18/a=009} {\bibfield  {journal}
  {\bibinfo  {journal} {J. Phys. A: Math. Gen.}\ }\textbf {\bibinfo {volume}
  {23}},\ \bibinfo {pages} {L987} (\bibinfo {year} {1990})}\BibitemShut
  {NoStop}%
\bibitem [{\citenamefont {Chame}\ and\ \citenamefont {Aar\~ao
  Reis}(2002)}]{Chame02}%
  \BibitemOpen
  \bibfield  {author} {\bibinfo {author} {\bibfnamefont {A.}~\bibnamefont
  {Chame}}\ and\ \bibinfo {author} {\bibfnamefont {F.~D.~A.}\ \bibnamefont
  {Aar\~ao Reis}},\ }\href {\doibase 10.1103/PhysRevE.66.051104} {\bibfield
  {journal} {\bibinfo  {journal} {Phys. Rev. E}\ }\textbf {\bibinfo {volume}
  {66}},\ \bibinfo {pages} {051104} (\bibinfo {year} {2002})}\BibitemShut
  {NoStop}%
\bibitem [{\citenamefont {Muraca}\ \emph {et~al.}(2004)\citenamefont {Muraca},
  \citenamefont {Braunstein},\ and\ \citenamefont {Buceta}}]{Muraca04}%
  \BibitemOpen
  \bibfield  {author} {\bibinfo {author} {\bibfnamefont {D.}~\bibnamefont
  {Muraca}}, \bibinfo {author} {\bibfnamefont {L.~A.}\ \bibnamefont
  {Braunstein}}, \ and\ \bibinfo {author} {\bibfnamefont {R.~C.}\ \bibnamefont
  {Buceta}},\ }\href {\doibase 10.1103/PhysRevE.69.065103} {\bibfield
  {journal} {\bibinfo  {journal} {Phys. Rev. E}\ }\textbf {\bibinfo {volume}
  {69}},\ \bibinfo {pages} {065103(R)} (\bibinfo {year} {2004})}\BibitemShut
  {NoStop}%
\bibitem [{\citenamefont {Silveira}\ and\ \citenamefont {Aar\~ao
  Reis}(2012)}]{Silveira12}%
  \BibitemOpen
  \bibfield  {author} {\bibinfo {author} {\bibfnamefont {F.~A.}\ \bibnamefont
  {Silveira}}\ and\ \bibinfo {author} {\bibfnamefont {F.~D.~A.}\ \bibnamefont
  {Aar\~ao Reis}},\ }\href {\doibase 10.1103/PhysRevE.85.011601} {\bibfield
  {journal} {\bibinfo  {journal} {Phys. Rev. E}\ }\textbf {\bibinfo {volume}
  {85}},\ \bibinfo {pages} {011601} (\bibinfo {year} {2012})}\BibitemShut
  {NoStop}%
\bibitem [{\citenamefont {Torres}\ and\ \citenamefont
  {Buceta}(2017)}]{Torres17}%
  \BibitemOpen
  \bibfield  {author} {\bibinfo {author} {\bibfnamefont {M.}~\bibnamefont
  {Torres}}\ and\ \bibinfo {author} {\bibfnamefont {R.}~\bibnamefont
  {Buceta}},\ }\href {https://arxiv.org/abs/1711.09652} {\bibfield  {journal}
  {\bibinfo  {journal} {arXiv:1711.09652}\ } (\bibinfo {year}
  {2017})}\BibitemShut {NoStop}%
\bibitem [{\citenamefont {Oliveira}\ \emph {et~al.}(2012)\citenamefont
  {Oliveira}, \citenamefont {Ferreira},\ and\ \citenamefont
  {Alves}}]{Oliveira12PRE}%
  \BibitemOpen
  \bibfield  {author} {\bibinfo {author} {\bibfnamefont {T.~J.}\ \bibnamefont
  {Oliveira}}, \bibinfo {author} {\bibfnamefont {S.~C.}\ \bibnamefont
  {Ferreira}}, \ and\ \bibinfo {author} {\bibfnamefont {S.~G.}\ \bibnamefont
  {Alves}},\ }\href {\doibase 10.1103/PhysRevE.85.010601} {\bibfield  {journal}
  {\bibinfo  {journal} {Phys. Rev. E}\ }\textbf {\bibinfo {volume} {85}},\
  \bibinfo {pages} {010601(R)} (\bibinfo {year} {2012})}\BibitemShut {NoStop}%
\bibitem [{\citenamefont {Ferrari}\ and\ \citenamefont
  {Frings}(2011)}]{Ferrari11}%
  \BibitemOpen
  \bibfield  {author} {\bibinfo {author} {\bibfnamefont {P.~L.}\ \bibnamefont
  {Ferrari}}\ and\ \bibinfo {author} {\bibfnamefont {R.}~\bibnamefont
  {Frings}},\ }\href {https://doi.org/10.1007/s10955-011-0318-4} {\bibfield
  {journal} {\bibinfo  {journal} {J. Stat. Phys.}\ }\textbf {\bibinfo {volume}
  {144}},\ \bibinfo {pages} {1123} (\bibinfo {year} {2011})}\BibitemShut
  {NoStop}%
\bibitem [{\citenamefont {Takeuchi}(2012)}]{Takeuchi12}%
  \BibitemOpen
  \bibfield  {author} {\bibinfo {author} {\bibfnamefont {K.~A.}\ \bibnamefont
  {Takeuchi}},\ }\href {http://stacks.iop.org/1742-5468/2012/i=05/a=P05007}
  {\bibfield  {journal} {\bibinfo  {journal} {J. Stat. Mech.: Theory Exp.}\
  }\textbf {\bibinfo {volume} {2012}},\ \bibinfo {pages} {P05007} (\bibinfo
  {year} {2012})}\BibitemShut {NoStop}%
\bibitem [{\citenamefont {Calabrese}\ and\ \citenamefont
  {Le~Doussal}(2011)}]{Calabrese11}%
  \BibitemOpen
  \bibfield  {author} {\bibinfo {author} {\bibfnamefont {P.}~\bibnamefont
  {Calabrese}}\ and\ \bibinfo {author} {\bibfnamefont {P.}~\bibnamefont
  {Le~Doussal}},\ }\href {\doibase 10.1103/PhysRevLett.106.250603} {\bibfield
  {journal} {\bibinfo  {journal} {Phys. Rev. Lett.}\ }\textbf {\bibinfo
  {volume} {106}},\ \bibinfo {pages} {250603} (\bibinfo {year}
  {2011})}\BibitemShut {NoStop}%
\bibitem [{\citenamefont {Imbrie}\ and\ \citenamefont
  {Spencer}(1988)}]{Imbrie88}%
  \BibitemOpen
  \bibfield  {author} {\bibinfo {author} {\bibfnamefont {J.~Z.}\ \bibnamefont
  {Imbrie}}\ and\ \bibinfo {author} {\bibfnamefont {T.}~\bibnamefont
  {Spencer}},\ }\href {https://doi.org/10.1007/BF01019720} {\bibfield
  {journal} {\bibinfo  {journal} {J. Stat. Phys.}\ }\textbf {\bibinfo {volume}
  {52}},\ \bibinfo {pages} {609} (\bibinfo {year} {1988})}\BibitemShut
  {NoStop}%
\bibitem [{\citenamefont {Gomes}\ \emph {et~al.}(2019)\citenamefont {Gomes},
  \citenamefont {Penna},\ and\ \citenamefont {Oliveira}}]{Gomes19}%
  \BibitemOpen
  \bibfield  {author} {\bibinfo {author} {\bibfnamefont {W.~P.}\ \bibnamefont
  {Gomes}}, \bibinfo {author} {\bibfnamefont {A.~L.~A.}\ \bibnamefont {Penna}},
  \ and\ \bibinfo {author} {\bibfnamefont {F.~A.}\ \bibnamefont {Oliveira}},\
  }\href {\doibase 10.1103/PhysRevE.100.020101} {\bibfield  {journal} {\bibinfo
   {journal} {Phys. Rev. E}\ }\textbf {\bibinfo {volume} {100}},\ \bibinfo
  {pages} {020101(R)} (\bibinfo {year} {2019})}\BibitemShut {NoStop}%
\end{thebibliography}%
 
\end{document}